\documentclass[10pt,journal,compsoc]{IEEEtran}
\ifCLASSOPTIONcompsoc
  \usepackage[nocompress]{cite}
\else
  \usepackage{cite}
\fi
\ifCLASSINFOpdf
\else
\fi
\DeclareMathAlphabet{\mathpzc}{OT1}{pzc}{m}{it}
\usepackage{subfigure}
\usepackage{graphicx}
\usepackage{caption}
\usepackage[bitstream-charter]{mathdesign}
\usepackage[T1]{fontenc}
\usepackage{threeparttable}
\usepackage{algorithmic}
\usepackage[ruled,vlined,linesnumbered]{algorithm2e}
\usepackage{multirow}
\usepackage{booktabs}

\begin{document}

\title{Heterogeneous Social Event Detection via Hyperbolic Graph Representations}

\author{Zitai~Qiu,
        Jia~Wu,~\IEEEmembership{Senior Member,~IEEE,}
        Jian~Yang,
        Xing~Su,
        Charu C. Aggarwal,~\IEEEmembership{Fellow,~IEEE}
\IEEEcompsocitemizethanks{\IEEEcompsocthanksitem Z. Qiu, J. Wu, J. Yang and Xing Su are with the School of Computing, Macquarie University, Sydney, NSW 2109, Australia.\protect\\
E-mail: \{zitai.qiu@students., jia.wu, jian.yang, xing.su2@hdr.\}mq.edu.au
\IEEEcompsocthanksitem C. Aggarwal is with IBM T. J. Watson Research Center, New York, USA.
E-mail: charu@us.ibm.com.
}
\thanks{Manuscript received April 19, 2005; revised August 26, 2015.}}

\markboth{Journal of \LaTeX\ Class Files,~Vol.~14, No.~8, August~2015}%
{Shell \MakeLowercase{\textit{et al.}}: Bare Demo of IEEEtran.cls for Computer Society Journals}

\IEEEtitleabstractindextext{%
\renewcommand{\raggedright}{\leftskip=0pt \rightskip=0pt plus 0cm}
\raggedright
\begin{abstract}
Social events reflect the dynamics of society and, here, natural disasters and emergencies receive significant attention. The timely detection of these events can provide organisations and individuals with valuable information to reduce or avoid losses. However, due to the complex heterogeneities of the content and structure of social media, existing models can only learn limited information; large amounts of semantic and structural information are ignored. In addition, due to high labour costs, it is rare for social media datasets to include high-quality labels, which also makes it challenging for models to learn information from social media. In this study, we propose two hyperbolic graph representation-based methods for detecting social events from heterogeneous social media environments. For cases where a dataset has labels, we designed a \textbf{\underline{H}}yperbolic \textbf{\underline{S}}ocial \textbf{\underline{E}}vent \textbf{\underline{D}}etection (HSED) model that converts complex social information into a unified social message graph. This model addresses the heterogeneity of social media, and, with this graph, the information in social media can be used to capture structural information based on the properties of hyperbolic space. For cases where the dataset is unlabelled, we designed an \textbf{\underline{U}}nsupervised \textbf{\underline{H}}yperbolic \textbf{\underline{S}}ocial \textbf{\underline{E}}vent \textbf{\underline{D}}etection (UHSED). This model is based on the HSED model but includes graph contrastive learning to make it work in unlabelled scenarios. Extensive experiments demonstrate the superiority of the proposed approaches.
\end{abstract}

}

\maketitle

\IEEEdisplaynontitleabstractindextext

%
\IEEEpeerreviewmaketitle

\IEEEraisesectionheading{\section{Introduction}\label{sec:introduction}}

\IEEEPARstart{E}{ven} are happenings in a community. They can be innocuous, such as a fair or town meeting, or they can have a significant harmful impact, such as a natural disaster or state of emergency \cite{9927311}. Detecting these events of harmful impact is crucial because early detection can help organisations and individuals make timely responses to avoid danger and loss \cite{cao2021knowledge,9664363}. The rapid development of social media has gradually replaced traditional TV and newspapers as a tool for people to obtain information about events. The data shows that, by February 2021, Facebook and Twitter had exceeded 2.8 billion monthly active users, and total active users in these social networks exceeded 390 million across more than 200 different countries \cite{afyouni2022multi}. Hence, many are defining events posted on social media, such as Facebook or Twitter, as social events \cite{huang2021survey}.  And, for this reason, researchers have gradually shifted the focus of event detection to social event detection \cite{ren2022evidential, tijare2021survey,fedoryszak2019real, ren2022known}.

\begin{figure}[htbp]
    \centering
    \includegraphics[width=0.4\textwidth]{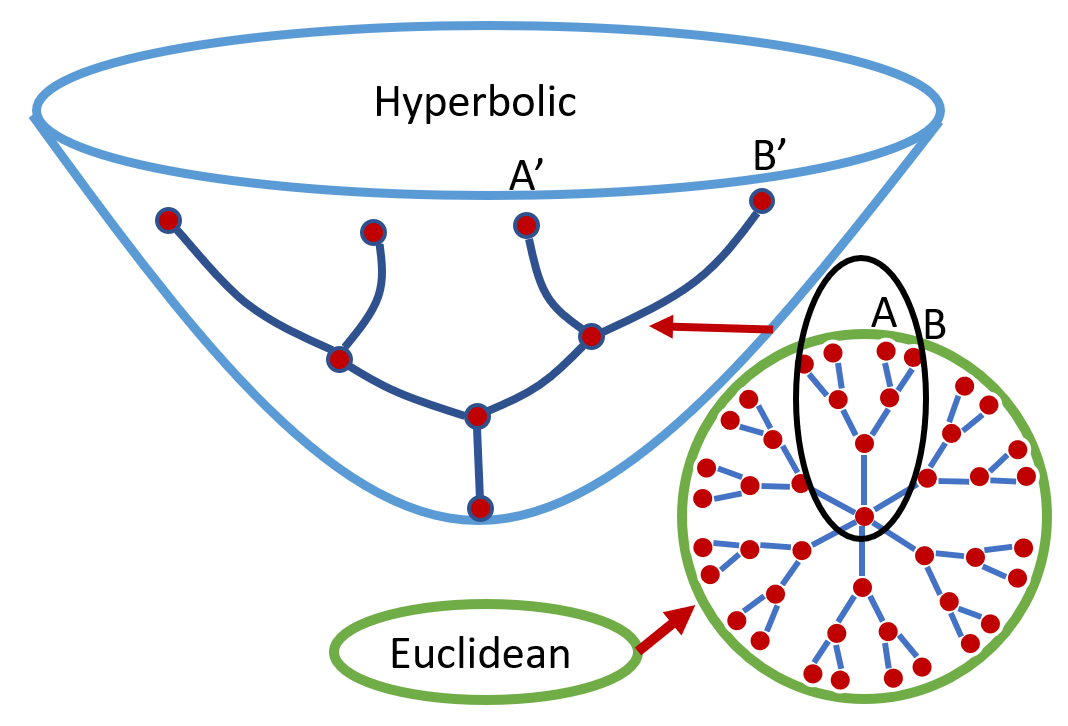}
    \caption[Euclidean and hyperbolic space representations]{Within a tree-like structure data, the distance between node $A$ and node $B$ is difficult to calculate in Euclidean space, but the distance between node $A'$ and node $B'$ is easy to calculate. Here, $A'$ and $B'$ are the projections of nodes A and B on hyperbolic space.}
    \label{EH_representation}
\end{figure}

The key idea of social event detection is text classification or clustering – that is, extracting relevant information from social media to represent a specific event  \cite{huang2021survey,cao2021knowledge}. However, compared to detection with traditional news, there are several challenges in social event detection shown as follow \cite{tijare2021survey,afyouni2022multi,RenSocial2021,fedoryszak2019real}: \textit{1)} The text describing social events are often short texts, not written by people, that frequently contain abbreviations, misspellings, and emojis.  \textit{2)} Messages on social media include a large amount of heterogeneous content: users, times, places, entities, and so on. \textit{3)} The spread of social events is mainly caused by user mentions and retweets, which leads to the hierarchical and heterogeneous nature of social networks. \textit{4)} Social media data is generated continuously and dynamically, which is more difficult to analyse and use than offline datasets.

Traditional social event detection techniques like LDA \cite{blei2003latent} mainly address the first challenge of social media. Generally, they use statistical methods to calculate the co-occurrence of words relating to a topic in short pieces of text \cite{huang2021survey}. However, word co-occurrence in short texts tends to be very sparse, which has hindered the ongoing development of topic detection models. Inspired by human understanding of short texts, Li and others \cite{li2016topic} proposed a GPU-DMM model. GPU-DMM includes a pre-trained Word2Vec \cite{mikolov2013efficient} model that provides semantic background information, which enhances the model’s performance. However, although traditional social event detection technology has achieved good results, it can be difficult for the models to capture word co-occurrence information because short texts do not include many words. Hence, in general, traditional models only focus on the semantic information of the message; they ignore the relevance of semantic and structural information discussed in  \textit{2)} and \textit{3)}.

The key idea for addressing challenges \textit{2)} and \textit{3)} – heterogeneity in content and structure – is how to learn helpful information from heterogeneous social media environments. We can model social media as a heterogeneous information network (HIN) \cite{sun2013mining, huang2020heterogeneous, wei2022heterogeneous}. Compared with homogeneous information networks, HIN contains multiple types of nodes and edges, which can integrate more information. However, due to heterogeneous complex content and structure of information networks, researchers cannot directly use them in traditional models. This is not only because of the need to incorporate heterogeneous structural (graph) information consisting of multiple nodes and edges but also because of the need to consider the heterogeneous attributes or content associated with each node. To address challenges \textit{2)} and \textit{3)}, PP-GCN \cite{peng2019fine} focuses on the heterogeneity of social media. It uses meta-path \cite{sun2011pathsim} to express the knowledge in heterogeneous information networks composed of social media. This better helps the model to learn the semantic and structural information in the network. 
Furthermore, to address challenge \textit{4)} – the dynamic nature of social media data – one state-of-the-art model, KPGNN \cite{cao2021knowledge}, addresses all the challenges faced in social event detection. However, the focus of this model is on how the model can be applied to a constantly changing social media stream. Its ability to learn from heterogeneous information networks is not significantly improved compared to previous models. It still only learns limited social media information, ignoring rich semantic and structural information.  Therefore, we believe that the premise of solving challenge \textit{4)} is to learn more helpful information from the heterogeneous social media environments. This will require overcoming the shortfalls of existing models, particularly since they can only learn small amounts of knowledge from social media.

Meta-paths are currently the most common technique for researchers to learn information from heterogeneous information networks. Of the currently available methods, very classic models such as HAN \cite{wang2019heterogeneous} and MAGNN \cite{fu2020magnn} have achieved competitive performance. However, the fatal disadvantage of meta-paths is that experienced personnel are required to design them  \cite{zhong2020reinforcement}. For datasets with large amounts of data, artificially creating meta-paths is very time-consuming and challenging. Moreover, experiments \cite{lv2021we} have shown that the meta-path method may have no practical effect, and it is not as good as the GCN \cite{kipf2016semi} or GAT \cite{velivckovic2017graph} for modelling isomorphic networks, assuming the appropriate parameter settings are being used. Additionally, because of the vast amounts of social media data – i.e.,  graphs with tens of thousands of points and complex relationships – it would be incredibly challenging and time-consuming to design meta-paths manually. Based on these shortcomings, we do not find the meta-path approach to be suitable for heterogeneous social media networks. As such, learning more information from heterogeneous information networks without using meta-paths remains a challenge.

It is worth noting that all the social event detection models mentioned above ignore a very critical factor: the data structure of a large-scale social network. The data generated by social media is tree-like or hierarchical in structure  \cite{adcock2013tree}. The most significant characteristic of data with this type of structure is that it grows exponentially. However, due to the polynomial growth of Euclidean space, existing models for mapping social media data to Euclidean space do not express or capture tree-like structures well \cite{yang2022dual,cannon1997hyperbolic}. As shown in Fig. \ref{EH_representation}, in Euclidean space, when the tree-like structure data gradually increases, the distance between the leaves becomes very close, which means that machine learning models cannot distinguish nodes at the leaf position. Inspired by geometric graph mining in hyperbolic space in recent years, we argue that because hyperbolic space also has an exponential growth property, it is more capable of capturing social media data than Euclidean space. However, the only existing hyperbolic space models are based on homogeneous information networks. A hyperbolic space model needs to be created for heterogeneous information networks.

To tackle the above challenges, we propose a social event detection model based on hyperbolic space representation. We call it \textbf{\underline{H}}yperbolic \textbf{\underline{S}}ocial \textbf{\underline{E}}vent \textbf{\underline{D}}etection (HSED). HSED incorporates Word2Vec to unify complex social information components into a homogeneous message graph to solve challenges  \textit{1)} and \textit{2)} while also further promoting the use of information. To solve challenge \textit{3)} and better express the structural information of the social media, HSED projects the homogeneous message graph generated in the previous step onto hyperbolic space. However, as mentioned, it can be very costly to label data. Hence, we also developed a variant of HSED called \textbf{\underline{U}}nsupervised \textbf{\underline{H}}yperbolic \textbf{\underline{S}}ocial \textbf{\underline{E}}vent \textbf{\underline{D}}etection (UHSED), which includes graph contrastive learning. Thus, between the two models, our approach can be applied to multiple scenarios.

The key contributions of this study are as follows:
\begin{itemize}
\item This approach is the first to apply hyperbolic space to social event detection as a better way of representing social media data than Euclidean space.

\item We designed a supervised model called HSED that converts heterogeneous social information into a homogeneous message graph and employs hyperbolic space as a way to leverage social media data.

\item For cases where the social media data is unlabelled, we designed an unsupervised model called UHSED, which is based on graph contrastive learning. Between these two models, our approach works with most social media datasets.

\item Experiments demonstrate the competitive performance of our models and the superiority of hyperbolic space for tree-structured data.

\end{itemize} 


\section{Related Work}
\subsection{Topic Detection-based Social Event Detection Models}
Today, social networks are ubiquitous. Social networks are a type of information network containing a great deal of rich content and a multitude of relationships, commonly modelled as a heterogeneous information network (HIN)  \cite{shi2016survey}. However, most existing models for social event detection are based on homogeneous information networks  \cite{sun2013mining}, i.e., networks that only contain nodes and edges of the same type. These models ignore the heterogeneity of nodes and edges \cite{shi2016survey}, which can easily cause severe data loss. Therefore, more and more researchers in social event detection are focussing on learning from heterogeneous information networks where semantic and structural information can be captured in addition to counting the number of co-occurring words.

Topic detection-based models and HIN-based social event detection models are the main offline social event detection models
\cite{huang2021survey,afyouni2022multi}. Social media detection models based on topic detection, like LDA \cite{blei2003latent}, focus on analysing short texts. Here, the topics in short pieces of text are captured by calculating the number of co-occurring words via mathematical statistics. However, this method has a limitation in that word co-occurrence in short pieces of text can be very sparse. And, when there are not enough co-occurring words, it seriously affects the model’s performance \cite{huang2021survey}. Therefore,  inspired by the idea that people reading prose understand more than just the words given but also the meaning behind the words  \cite{huang2021survey},  i.e., the semantic information, researchers have turned to semantics to extract more knowledge from these brief parcels of data.

Among the topic detection models, Word2Vec \cite{mikolov2013efficient} is commonly used to find semantically similar topic words in short texts through word vectors. For example, GPU-DMM \cite{li2016topic} applies Word2Vec based on LDA and DMM \cite{ma2014bayesian} to provide the model with background information on word semantics, which improves the model’s performance. However, although Word2Vec does provide some background information to the model, there is still too little keyword information in short text to be able to improve the model’s performance. Additional helpful information needs to be learned from more aspects of the data. For example,  SGNS \cite{shi2018short} starts with the relationship between words. It learns the relationship between words, captures the semantic relationship of the context of words, and, in so doing, makes up for the problem of keyword sparsity in short texts. However, topic detection models always ignore heterogeneity in social media networks and, hence, they ignore a great deal of semantic and structural information.

\begin{figure}[htbp]
    \centering
    \includegraphics[width=0.5\textwidth]{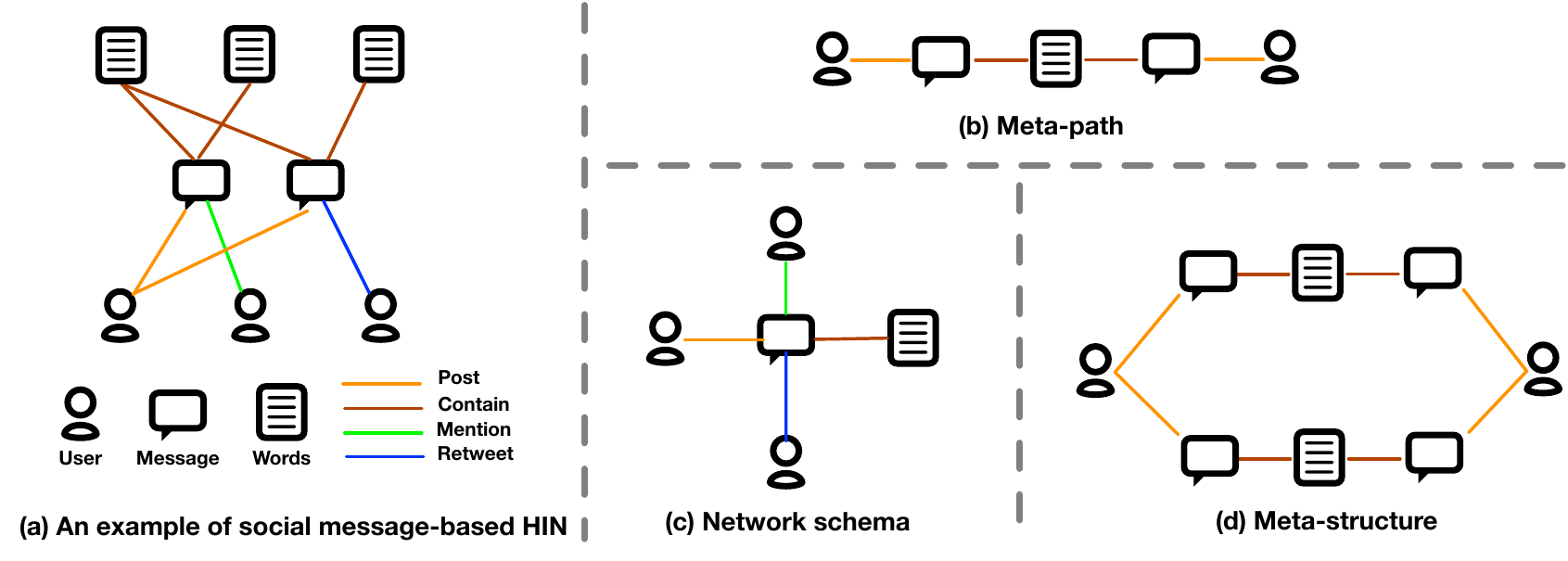}
    \caption[An illustrative example of a HIN]{An illustrative example of a heterogeneous message graph: (a) A social message network containing three types of notes (e.g., user, message, and words) and four types of links (e.g., post, contain, mention, retweet). (b) The network schema of a social message network. (c) An example of a meta-path in a social message network. (d) An example of a meta-structure.}
    \label{fg:structure_semantic}
\end{figure}

\subsection{Heterogeneous Information Networks-based Social Event Detection Models}
In the social event detection models based on heterogeneous information networks, researchers model social messages as a heterogeneous information network. The most popular method of modelling is the meta-path \cite{sun2011pathsim}. Through these constructs, the models capture both semantic and structural information simultaneously, as shown in Fig. \ref{fg:structure_semantic}, resulting in data we call structural-semantic information. However, these meta-paths must be designed by experts. The limitation of the meta-path is that it sometimes cannot find the same points in different meta-paths \cite{huang2016meta}. Therefore, the meta-structure \cite{huang2016meta} can contain more information based on the meta-path, as shown in Fig. \ref{fg:structure_semantic}(d). The social event detection model PP-GCN \cite{peng2019fine} includes a meta-schema based on meta-structure, which describes the semantic associations between social events. PP-GCN also incorporates knowledge-based meta-paths where the similarity between different meta-paths is calculated to distinguish between different events in the network.  However, both the meta-paths and the meta-structures are artificially set, and with a substantial amount of data, the candidate meta-paths grow exponentially \cite{huang2016meta}. Consequently, artificially designing accurate and meaningful meta-paths and meta-structures for large-scale datasets is very difficult and time-consuming.

\begin{figure}[htbp]
    \centering
    \includegraphics[width=0.49\textwidth]{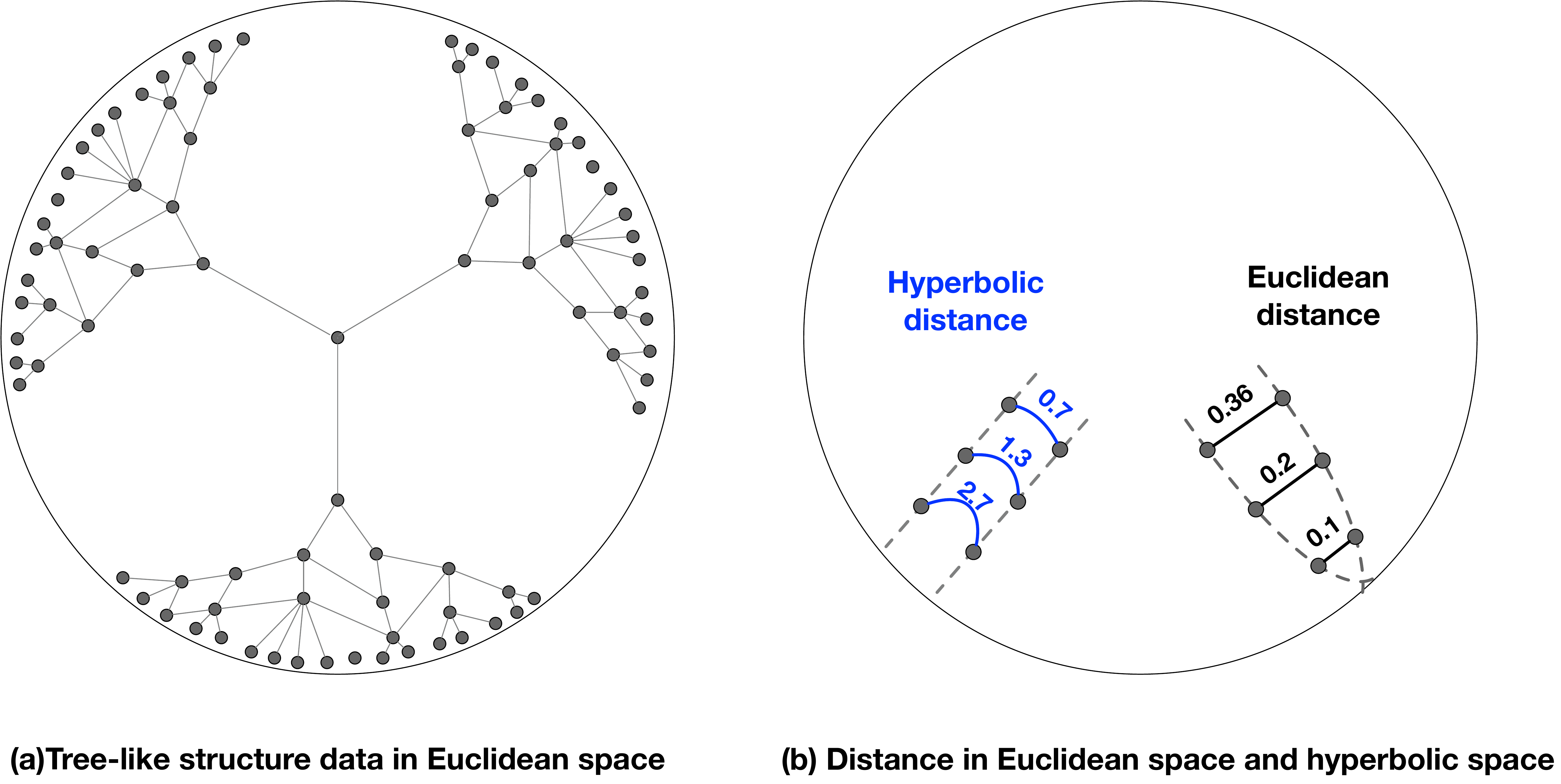}
    \caption[An example of tree-like structure data]{An example of tree-like structure data. (a) An example of tree-structured data expanding exponentially in Euclidean space. (b) Comparison of the distance between two nodes in hyperbolic space (blue) and Euclidean space (black).}
    \label{fg:HE_distance}
\end{figure}
Notably, meta-paths and meta-structures capture a kind of structure-based semantic information, local information.  The social network data is tree-structured \cite{ganea2018hyperbolic}, as shown in Fig. \ref{fg:HE_distance}(a). Models using meta-paths or meta-structures may not capture the relationship between root and leaf. In addition, existing models embed nodes onto Euclidean space to capture features. When the data grows faster than the Euclidean space can expand to, the leaf parts of the tree structure will become very close, and the distance between them is infinitely based on zero. This makes it difficult for the model to distinguish the difference between them, as shown in Fig. \ref{fg:HE_distance}(b). In general, existing social event detection models cannot make good use of the distances between structures in heterogeneous social information networks due to the above limitation of Euclidean space.

\subsection{Hyperbolic Representation Learning}
Currently, most machine learning models choose Euclidean space as the main space for embedding learning due to its convenient distance computation and vector structure  \cite{ganea2018hyperbolic}. Although Euclidean spatial embeddings have been successful, most real-world graphs, such as social networks, exhibit tree-like structures  \cite{peng2021hyperbolic}. Recent studies have shown that the Euclidean space with polynomial growth is not good at providing meaningful geometric representation for tree-structured data and that it can lead to significant distortions after embedding \cite{liu2019hyperbolic, chami2019hyperbolic}. Unlike Euclidean space, growth in hyperbolic space is exponential, just like growth with tree-structured data. Thus, hyperbolic space should be able to provide a more powerful geometric representation of tree-structured data and, in so doing, reduce the distortion after embedding  \cite{ganea2018hyperbolic, fu2021ace}. However, there are no basic statistical algorithms, like vector addition and matrix multiplication, for hyperbolic space  \cite{ganea2018hyperbolic, peng2021hyperbolic}. Therefore, many machine learning algorithms, such as MLP and RNN, cannot be used in hyperbolic space.

There are, however, some exceptions. HNN \cite{ganea2018hyperbolic}, for example, uses a simple neural network, such as an MLP or an RNN, with hyperbolic space. However, this framework only considers graph structures; it ignores the rich node features. Therefore, on the basis of HNN, HGCN \cite{chami2019hyperbolic} applied a more complex GCN model to
hyperbolic space. Its main contribution is how to aggregate neighbour information in hyperbolic space. Combining the advantages of the above hyperbolic space models, we can see that hyperbolic space can capture the distance information between nodes better than Euclidean space. However, all these models only work with homogeneous information networks. This leads to our inability to directly apply them to heterogeneous social media. As mentioned above, hyperbolic space is a non-vector space. Although there are many methods for learning heterogeneous information network representation in Euclidean space, these methods cannot be directly used on hyperbolic space due to the problem of vector calculation. We need to redesign methods for learning representations of heterogeneous information networks in hyperbolic spaces to address the above challenges, but this is still an open problem.

\subsection{Graph Contrastive Learning} \label{GCL}

Graph Contrastive Learning (GCL) is currently the most popular technique for unsupervised representation learning. DGI \cite{velickovic2019deep} employs Deep Infomax \cite{hjelm2018learning} for graph learning and helps the model to learn by maximising the mutual information between the local structures and the global contexts as a pre-task.  Based on DGI, GCC \cite{qiu2020gcc} compares different sub-graphs extracted from the original graph. GCC mainly learns by maximising the mutual information between sub-graphs as a pre-task. Unlike GCC, GraphCL \cite{you2020graph} forms a graph as a negative sample through graph data enhancement, which compares the original and enhanced graphs to maximise the mutual information between them.

\section{Preliminaries}

Our two models aim to capture the distances between nodes in tree-like social media data. To this end, the models embed features into hyperbolic space, which is a better way of representing this type of data. To better understand our methods, Section \ref{HIN} introduces some preliminary concepts associated with heterogeneous information networks; Section  \ref{MHS} deals with hyperbolic space; and Section \ref{HGR} deals with hyperbolic graph representations.

\subsection{Heterogeneous Information Network} \label{HIN}
We defined a graph $\mathpzc{G} = (V, E)$ as an information network. $V$ denotes a set of nodes and  $E$ denotes a set of edges. Given a set of node types $T_v$ and a set of edge types $T_E$. When $|T_v| > 1$ or  $|T_E|> 1$, the information network is \textbf{heterogeneous}; otherwise, it is a \textbf{homogeneous}. 

Most machine-learning models embed features into Euclidean space to capture the information from the information network; however, as mentioned, Euclidean space cannot represent tree-like structure data well. Thus, in this study, we have selected hyperbolic space as the embedding space. The following sections outline the details of hyperbolic space.

\subsection{Models of Hyperbolic Space} \label{MHS}
In the past, researchers have  proposed different models of hyperbolic space based on different usage scenarios, like the Poincaré ball model $\mathbb{P}$, the Interior of the disk model $\mathbb{I}$, the Hyperboloid model $\mathbb{H}$ and the Jemisphere model $\mathbb{J}$ \cite{cannon1997hyperbolic}. This thesis mainly focuses on the Poincaré ball and the hyperboloid models based on their properties for this study.
\subsubsection{The Poincaré Ball Model}
The Poincaré ball model can be adjusted via gradient-based optimisation \cite{nickel2017poincare}. $\mathbb{P} ^{d, \mathpzc{K}}$ denotes the Poincaré ball model with a constant negative curvature of $-\mathpzc{K} (\mathpzc{K} > 0)$ in $d$-dimensions:
\begin{equation}
    \mathbb{P}^{d, \mathpzc{K}} := \{ 
    x \in \mathbb{R} ^{d} : \mathpzc{K}\|x\|^2 < 1
    \}.
    \label{eq:poincare}
\end{equation}
Given two nodes $a,b$ in the Poincare ball model $\mathbb{P} ^{d,\mathpzc{K}}$, the induced distance between them can be calculated as:
\begin{equation}
    d_\mathbb{P} ^ \mathpzc{K} (a,b) = \frac{1}{arcosh \left(1 +  \frac{2\|a - b\|^2}{(1-\|a\|^2)(1-\|b\|^2)} \right)},
    \label{eq:P_distance}
\end{equation}
where $\mathpzc{K}=1$,  $\|\cdot\|$ means the Euclidean norm and $arcosh(\cdot)$ means the arc cosine function in hyperbolic space.

\subsubsection{The Hyperboloid Model} 
The properties of the hyperboloid model are simplicity and numerical stability \cite{nickel2018learning}. The hyperboloid model is also called the Minkowski model because it can be defined by Minkowski's inner product \cite{chami2019hyperbolic}, expressed as follow:
\begin{equation}
    \langle .,.\rangle _ \mathpzc{M} : \mathbb{R} ^{d + 1} \times \mathbb{R} ^{d + 1} \rightarrow \mathbb{R}
    \label{eq:Minkowshi},
\end{equation}
\begin{equation}
    \langle x, y \rangle _ \mathpzc{M} := -x_0y_0 + x_1y_1 + \dots + x_n y_n.
\end{equation}
 $\mathbb{H} ^{d,\mathpzc{K}}$ denotes as the hyperboloid model with a constant negative curvature of $-/\mathpzc{K} (\mathpzc{K} > 0)$ in $d$-dimensions:
\begin{equation}
    \mathbb{H}^{d, \mathpzc{K}} := \{x \in \mathbb{R} ^{d + 1} : \langle x,x\rangle_ \mathpzc{M} = -\mathpzc{K}, x_0 > 0 \}.
    \label{eq:Hyboloid_model}
\end{equation}

Given two nodes $a,b$ in the hyperboloid model $\mathbb{H} ^{d,\mathpzc{K}}$, the induced distance between them can be calculated as:
\begin{equation}
    d^\mathpzc{K}_ \mathbb{H} (a,b) = \sqrt{\mathpzc{K}}\cdot arcosh(-\langle a,b\rangle_ \mathpzc{M}/\mathpzc{K}).
    \label{eq:hyperboloid_distance}
\end{equation}
\subsection{Hyperbolic Graph Representations} \label{HGR}
The challenge in adopting a graph representation learning method for hyperbolic space is that hyperbolic space is not vector space  \cite{zhang2021hyperbolic}. Specifically, the vector processing procedure of graph representation learning models in Euclidean space does not work in hyperbolic space. The solution to leveraging these models is to map the embeddings from hyperbolic space to Euclidean space. We can transform embeddings from Euclidean space to hyperbolic space via an exponential map. We also can transform embeddings from hyperbolic space to Euclidean space via a logarithmic map. Note that the mapping between these hyperbolic and Euclidean spaces is bijective \cite{chami2019hyperbolic}, which means that, for an infinite tangent space, there is a one-to-one corresponding point on the hyperbolic space.
Here, we explain this mapping mechanism for our two models.

\subsubsection{The Poincaré Ball Model Representations}
 $ T_o \mathbb{P}^{d, \mathpzc{K}}$ denotes the Euclidean (tangent) space centred at point $o$:

 Let $a \in \mathbb{P}^{d, \mathpzc{K}}$ and $ a' \in T_o \mathbb{P}^{d, \mathpzc{K}}$. The mapping from Euclidean space to hyperbolic space $exp^\mathpzc{K} _o : T_o \mathbb{P}^{d, \mathpzc{K}} \rightarrow \mathbb{P}^{d, \mathpzc{K}}$, and the mapping from hyperbolic space to Eudlicean space $log^\mathpzc{K} _o : \mathbb{P}^{d, \mathpzc{K}}  \rightarrow  T_x \mathbb{P}^{d, \mathpzc{K}}$. Thus we have:
 \begin{equation}
    exp^\mathpzc{K} _o (a') = tanh(\sqrt{\mathpzc{K}}\|a'\|) \frac{a'}{\sqrt{\mathpzc{K}}\|a'\|},
    \label{eq:P_exp}
\end{equation}

\begin{equation}
    log^\mathpzc{K} _o (a) = artanh(\sqrt{\mathpzc{K}}\|a\|) \frac{a}{\sqrt{\mathpzc{K}}\|a\|},
    \label{eq:P_log}
\end{equation}

\begin{equation}
    log^\mathpzc{K}_o(exp^\mathpzc{K} _o(a')) =a'.
\end{equation}

\subsubsection{The Hyperboloid Model Representations}
$ T_o \mathbb{H}^{d, \mathpzc{K}}$ denotes the tangent (Euclidean) space centred at point $o$:

 For $x, y \in \mathbb{H}^{d, \mathpzc{K}}$ and $v \in T_o \mathbb{H}^{d, \mathpzc{K}}$, where $v \neq 0$ and $x \neq y$, we have:
\begin{equation}
    exp^\mathpzc{K} _o (v) = cosh\left(\frac{\|v\|}{\sqrt{\mathpzc{K}}}\right)x + \sqrt{\mathpzc{K}} \cdot sinh\left(\frac{\|v\|}{\sqrt{\mathpzc{K}}}\right)\frac{v}{\|v\|},
    \label{eq:H_exp}
\end{equation}

\begin{equation}
    log^\mathpzc{K} _o (y) = d^\mathpzc{K}_\mathbb{H} (x,y) \frac{y+ \frac{1}{\mathpzc{K}}\langle x, y\rangle_ \mathpzc{M} x}{\|y+\frac{1}{\mathpzc{K}}\langle x,y\rangle _{\mathpzc{M}} x\|}.
    \label{eq:H_log}
\end{equation}

Therefore, based on the above transformation, the existing models in Euclidean space can be used to handle tasks in hyperbolic space.

\section{Methodology}

\subsection{HSED: A Supervised Model for Social Event Detection}

\subsubsection{Overview} \label{SOV}
 \begin{figure}[htbp]
    \centering
    \includegraphics[width=0.5\textwidth]{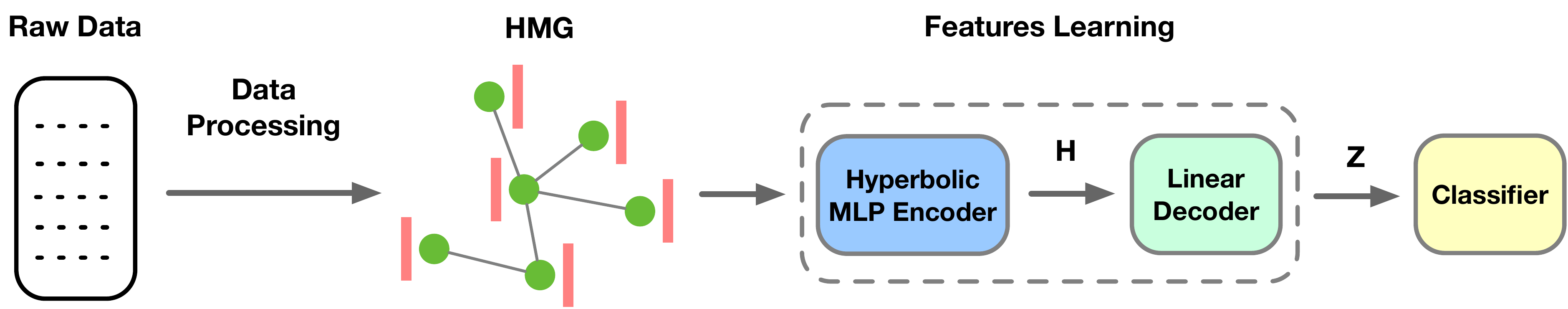}
    \caption[HSED framework]{The framework of the HSED model. “HMG” denotes a homogeneous message graph. “H” denotes hyperbolic embeddings after the hyperbolic encoder. “Z” denotes the final representations.}
    \label{fg:HSED}
\end{figure}

Motivated by recent advancements in hyperbolic graph representation learning, we propose Hyperbolic Social Event Detection (HSED) for heterogeneous social media
environments. The framework of our proposed model, shown in Fig. 
 \ref{fg:HSED}, consists of three main components:

\begin{enumerate}
    \item \textbf{Data processing}, which constructs a homogeneous information network  $\mathpzc{G}$  from the raw data while preserving the semantic and structural information.

    \item \textbf{Hyperbolic MLP encoder}, which embeds  $\mathpzc{G}$ into hyperbolic space to produce the feature embedding set  $\mathpzc{H}$.
    
    \item \textbf{Linear decoder}, which transfers  $\mathpzc{H}$ from hyperbolic space to Euclidean space through a  $log$ mapping function (Eq.\ref{eq:P_log} or Eq.\ref{eq:H_log}) to support downstream tasks. Here, the downstream task is node classification.

\end{enumerate}

\subsubsection{Data Processing} \label{DP}
Researchers cannot feed initial social media data directly into machine learning models. Therefore, the raw data needs to be processed to satisfy the conditions of the model. This is given that most hyperbolic graph representation learning models are based on homogeneous information networks. The challenge with data processing is transforming heterogeneous social networks into homogeneous ones while not ignoring the rich semantic and structural information that is also contained in the data.

Therefore, to fully leverage the information available, social messages are modelled into heterogeneous information networks (HINs) by the different types of entities in messages. However, not all entities in the message are essential. Inspirited by \cite{afyouni2022multi, cao2021knowledge}, the features in a message most strongly related to events are location information, temporal information, and semantic information.
\begin{figure}[htbp]
    \centering
    \includegraphics[width=0.5\textwidth]{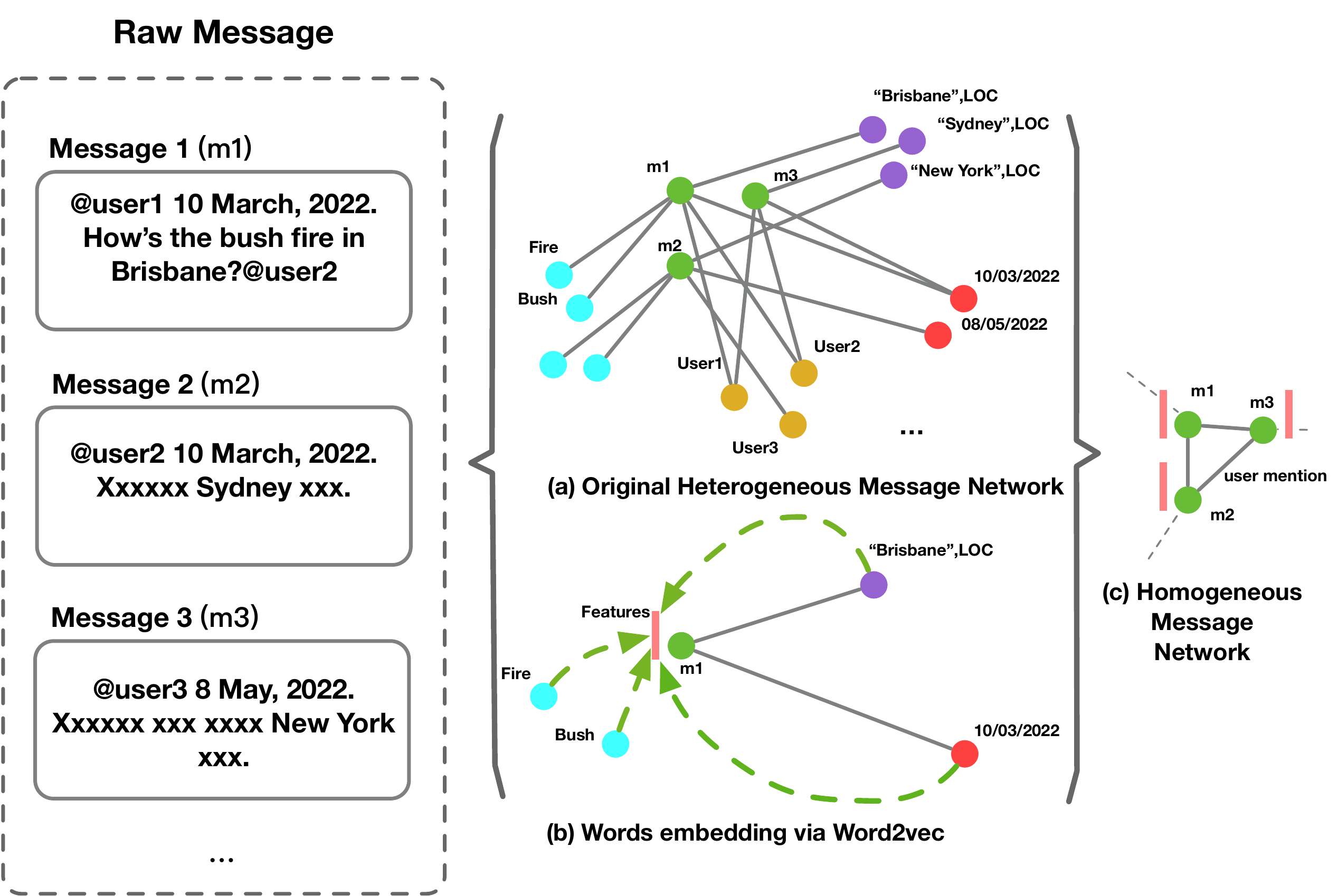}
    \caption[Social media data processing]{Social media data processing: (a) The original heterogeneous message network generated by the raw message. The different node colours denote different entities. (b) The process of learning message features via Word2vec. (c) The homogeneous message network generated from Steps (a) and (b).}
    \label{fg:data_processing}
\end{figure}
Our heterogeneous social information network was formed from messages on Twitter, as shown in Fig. \ref{fg:data_processing}(a). Given a message $m_i$ from a message set $M$ where $m_i \in M$, we treat the words in the message $m_i$ and the message itself as different entities. For example, “$m_1$” can be set as a message node. Based on the element in the message $m_1$,  “fire” and “bush” are selected as the word nodes; “Brisbane” is selected as the location node, and “10/03/2022” is set as the time node. The users mentioned in the messages or retweets are set as user nodes, e.g., “user1” and “user2”. Then edges are added between the message node and the other nodes. This process is repeated for all messages in the dataset.
Also, duplicate nodes are removed to build a heterogeneous message network that includes all node types.

Note, however, that is difficult to directly use a machine learning model with a heterogeneous information network. So the next step is to transform the heterogeneous message network into a homogeneous message network. We realise that in social media such as Twitter, user mentions and retweets play an important role in the spread of information. This is characteristic of networks and data with tree-like structures \cite{bian2020rumor}. Given this, we set message nodes as the only type of node in the homogeneous message graph, with user mentions and retweets as links between messages, as shown in Fig.  \ref{fg:data_processing}(c). To capture the semantic and time information, Word2Vec is used to learn the feature vectors of the semantic features of other nodes and to encode the timestamps into 2-dimensional vectors as time features. Then, the semantic and temporal are combined as message features, as shown in Fig.  \ref{fg:data_processing}(b). The message features denoted as $ X = \{x_{m_i} \in \mathbb{R}^d | m_i  \in M \} $, where $x_{m_i}$ is the features of message $m_i$. Thus, the homogeneous message graph can be expressed as $\mathpzc{G} = (X, A)$, where $A$ is the adjacency matrix of a homogeneous message network.

\subsubsection{Hyperbolic MLP Encoder and Linear Decoder} \label{H_MLP_Encoder}

To leverage the distance between nodes in tree-like data, hyperbolic space serves as the low-dimensional space for the embeddings needed to learn this information. Various hyperbolic representation learning models have been developed by researchers, from the
most straightforward model HNN 
 \cite{ganea2018hyperbolicnetwork} to some complex model like HGCN \cite{chami2019hyperbolic}, HGNN \cite{liu2019hyperbolic}, and HAT \cite{zhang2021hyperbolic}. 
 
 We note that a simple hyperbolic representation learning model has the same structure as a model designed for Euclidean space. The difference is that some of the processing is shifted to hyperbolic space. Inspired by HNN, our hyperbolic encoder is designed based on an MLP and a linear decoder. As previously mentioned, hyperbolic space is non-vector space. When an MLP is applied to hyperbolic space, traditional feature transformations will not work. Therefore, the hyperbolic embeddings need to be mapped in Euclidean space to compute embedding vectors via a weight matrix and a bias translation.

Feature transform for a feature set $X$ in Euclidean space is shown as:
\begin{equation}
    \mathpzc{E} = \sigma(WX+b), 
    \label{eq:MLP}
\end{equation}
where, $\mathpzc{E}$ is the embeddings in Euclidean space, with  the activation function $\sigma$,  weight matrix $W$ and bias translation $b$.

Let $W$ be a $N x N'$ weight matrix, we have:
\begin{equation}
    W \otimes X = exp_o^\mathpzc{K}(Wlog_o^\mathpzc{K}(X)),
    \label{eq:Weight matrix}
\end{equation}
for bias translation in hyperbolic space, we have:
\begin{equation}
    X \oplus b = exp_o^\mathpzc{K}(log_o^\mathpzc{K}(X) + b),
    \label{eq:bias translation}
\end{equation}
and the full feature transforms in hyperbolic space is:
\begin{equation}
    \mathpzc{H} = exp_o^\mathpzc{K}(\sigma(log_o^\mathpzc{K}(W \otimes X \oplus b))).
    \label{eq:bias hyperbolic feature transform}
\end{equation}

Given a graph $\mathpzc{G} = (X, A)$, and a hyperbolic MLP decoder $\mathpzc{m}(\cdot)$ based on the above hyperbolic feature transform, we have:
\begin{equation}
    \mathpzc{H}^{\mathbb{H}} = \mathpzc{m}(\mathpzc{G}),
    \label{eq:hyperbolic MLP embedding_H}
\end{equation}
or
\begin{equation}
    \mathpzc{H}^{\mathbb{P}} = \mathpzc{m}(\mathpzc{G}),
    \label{eq:hyperbolic MLP embedding_P}
\end{equation}
where, $\mathpzc{H}^{\mathbb{H}}$ means the hyperboloid model is used for the embedding, and $\mathpzc{H}^{\mathbb{P}}$means the Poincaré ball model is used for the embedding. The adjacency matrix $A$ is not used in the HSED model, but it is used in the unsupervised variant of the model.

Additionally, the embeddings in hyperbolic space cannot be directly used for downstream tasks. Rather, they need to be transformed into embeddings in Euclidean space. We chose node classification as our downstream task, and so chose a linear decoder, denoted as  $\mathpzc{d}(\cdot)$, to handle the transformation:
\begin{equation}
    \mathpzc{Z} = \mathpzc{d}(\mathpzc{H}^{\mathbb{H}}),
    \label{eq:linear_decoder_H}
\end{equation}
or
\begin{equation}
    \mathpzc{Z} = \mathpzc{d}(\mathpzc{H}^{\mathbb{P}}),
    \label{eq:linear_decoder_P}
\end{equation}
where $\mathpzc{Z}$ is the set of final representations for the node classification task.

\subsubsection{Classifier} \label{Sloss}
Our downstream task is more specifically multi-node classification. Hence, the HSED model includes softmax and cross-entropy loss.

Given a representation set $ \mathpzc{Z} = \{z_i \in \mathbb{R} | 1 \leq i \leq N \} $, where $N$ is the total number of messages, plus $\mathpzc{S}(\cdot)$as the softmax layer and  $\mathpzc{L}_{HSED}(\cdot)$ as the cross-entropy loss, the loss can be calculated to update the model’s parameters. The processes are as follows:

Let $z_i \in \mathpzc{Z}$, $n \in N$, we have:
\begin{equation}
    \mathpzc{S}(z_i) = \frac{e^{z_i}}{\sum_{j=1}^n e^{z_j}},
    \label{eq:softmax}
\end{equation}
\begin{equation}
    l_i' = \mathpzc{S}(z_i),
    \label{eq:predic label}
\end{equation}
\begin{equation}
    \mathpzc{L}_{HSED} = - \sum\limits_{i=1}^nl_ilogl_i',
    \label{eq:softmax}
\end{equation}
where $l_i$ is the true label and $l_i'$ is the predicted label. The pseudocode for the HSED model is given in Algorithm \ref{alg:HSED}.

\begin{algorithm}[!htbp]
\SetKwData{Left}{left}
\SetKwData{This}{this}
\SetKwData{Up}{up}
\SetKwFunction{Union}{Union}
\SetKwFunction{FindCompress}{FindCompress}
\SetKwInOut{Input}{Input}
\SetKwInOut{Output}{Output}
\caption{\textbf{HSED}: Hyperbolic Social Event Detection}
\label{alg:HSED}
\KwIn{A social message database $M=\{m_1,...,m_n\}$; A set of labels $\{l_0, l_1,...,l_n\}$;hyperbolic MLP encoder $\mathpzc{m}(\cdot)$; Linear decoder $\mathpzc{d}(\cdot)$ ;softmax layer $\mathpzc{S}(\cdot)$.}
\KwOut{Set of social event labels $\{l_0',l_1',...,l_n'\}$, where $n$ is the number of classes.}
\BlankLine
\For{$m_i \in M$}
{Homogeneous message network $\mathpzc{G} = (X, A) \leftarrow$ data processing (Section \ref{DP})}
\For{$x_i \in X$}
{Hyperbolic embeddings $h_i = \mathpzc{m}(x_i)$\\
Final representations $z_i = \mathpzc{d}(h_i)$\\
Predict label $l_i' = \mathpzc{S}(z_i)$\\
Cross-entropy loss $\mathpzc{L}_{HSED} = - \sum\limits_{i=1}^nl_ilogl_i'$\\
Update parameters}
\end{algorithm}

\subsection{UHSED: An Unsupervised Model for Social Event Detection}
\subsubsection{Overview} \label{UOV}

High-quality datasets in social media data are rare given the cost of labelling. Hence, we devised a second model called Unsupervised Hyperbolic Social Event Detection (UHSED) that includes a contrastive learning technique to make our approach compatible with unlabelled datasets. The framework of the UHSED model is shown in Fig.\ref{fg:UHSED_framework}. The framework is similar to the HSED model but it adds three more components:

\begin{figure}[htbp]
    \centering
    \includegraphics[width=0.5\textwidth]{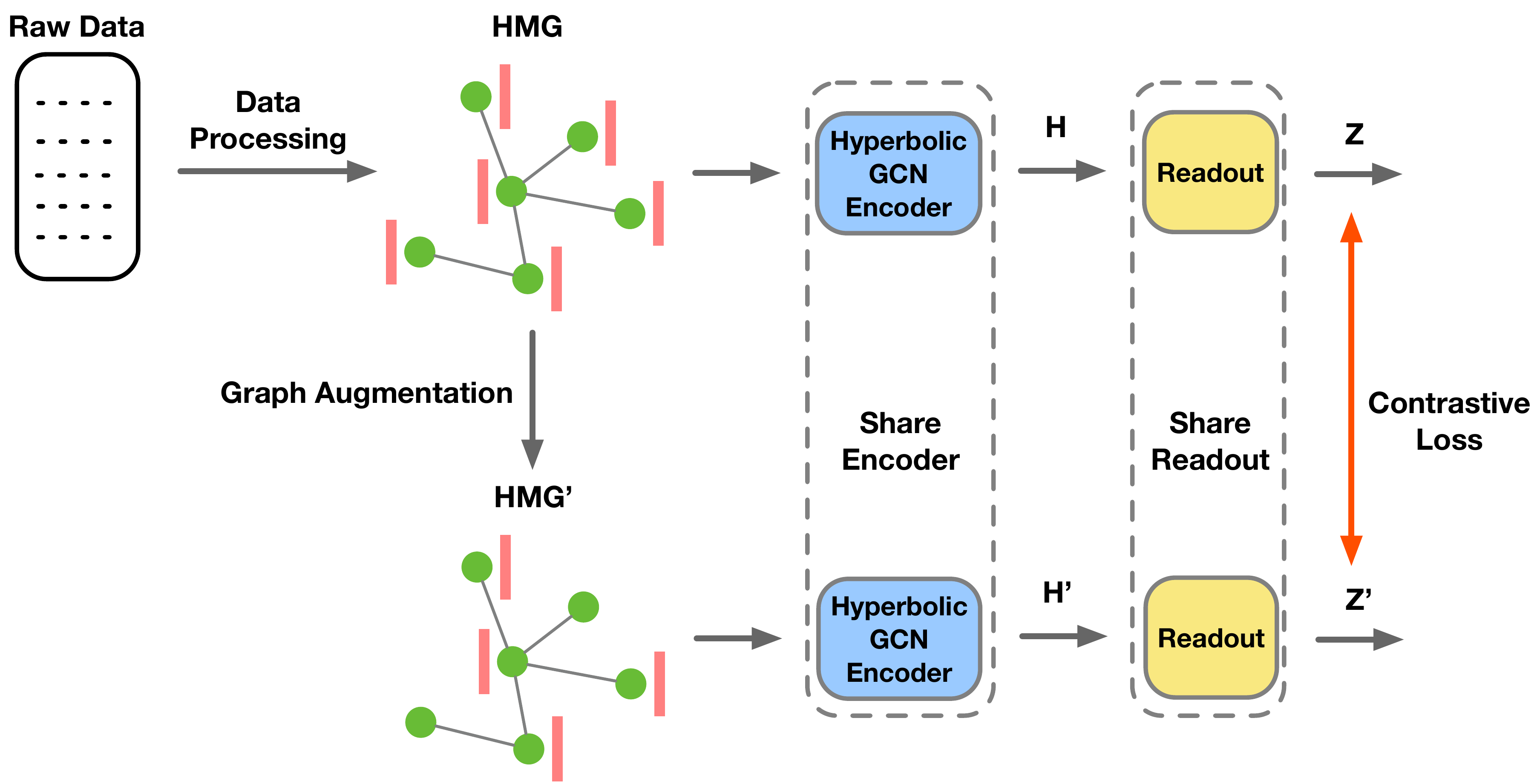}
    \caption[UHSED framework]{The framework of the UHSED model. “$HMG$” denotes a homogeneous message graph. “$HMG'$” denotes another homogeneous message graph after graph data augmentation. “$H, H'$” are the hyperbolic embeddings of the two HMGs after hyperbolic encoding. “$Z, Z'$” are the final representations.}
    \label{fg:UHSED_framework}
\end{figure}

\begin{enumerate}
    \item \textbf{Graph Data Augmentation}, which creates a negative sample and a graph  $\mathpzc{G'}$ of the original graph  $\mathpzc{G}$ for contrastive learning.
   
    \item \textbf{Hyperbolic GCN Encoder}, which embeds $\mathpzc{G}$ and $\mathpzc{G'}$ into the hyperbolic space to produce the patch embedding sets $\mathpzc{H}$ and $\mathpzc{H'}$.
    
    \item \textbf{Contrastive Loss}, which maximises the mutual information  $\mathpzc{H}$ and $\mathpzc{H'}$.
\end{enumerate}

\subsubsection{Graph Data Augmentation} \label{GDA}
In the data augmentation process, contrastive learning is performed in addition to the same data processing as used for the HSED model. Existing contrastive learning methods can be divided into CV-based contrastive learning, NLP-based contrastive learning, and graph-based contrastive learning depending on the application the model is being used for. Given that our data is graph-based, UHSED uses graph contrastive learning.

It is worth noting that data augmentation plays a vital role in contrastive learning. Researchers have proposed various data augmentation methods for different contrastive learning methods. As mentioned, the edges in social message networks hold vital structural information. Thus, only node-based data augmentation methods are considered in this study. The main node-based data augmentation methods are feature dripping, random masking, and feature corruption  \cite{you2020graph, velickovic2019deep}. Table \ref{tab:graph_data_a} provides an overview of these methods.
\begin{table}[!htb]
    \centering
    \caption[Overview of graph data augmentation methods]{Overview of graph data augmentation methods. }
    \scalebox{0.85}{
    \begin{tabular}{c|c|c}
    \toprule[1 pt]
    \textbf{Augmentation} & \textbf{Type} & \textbf{Description} \\ \midrule
    Feature dropping & Nodes &  Randomly select nodes and delete all their features  \\ \hline
    Random masking & Nodes & Random mask a certain percentage of features \\ \hline
    Feature corruption & Nodes & Disrupt the features corresponding to all nodes   \\
    \bottomrule[1pt]
    \end{tabular}
    }
    \label{tab:graph_data_a}
\end{table}

Given a homogeneous message graph $\mathpzc{G} = (X, A)$, a negative example is sampled through the graph data augmentation: $\mathpzc{G'} = (X', A')$. For all node-based data augmentations, $X \neq X'$ but $A = A'$.

\subsubsection{Hyperbolic GCN Encoder}
To fully leverage the power of graph contrastive learning, we developed a hyperbolic
GCN encoder to replace the hyperbolic MLP encoder in the HSED model. Compared to a hyperbolic MLP encoder, a hyperbolic GCN encoder based on a GCN model can
aggregate the information from neighbouring nodes to help the contrastive model capture mutual information. However, aggregation in hyperbolic space is different from aggregation in Euclidean space.

An aggregation in Euclidean space at layer $\mathpzc{l}$ would begin with
a feature transform:

\begin{equation}
    e_i^\mathpzc{l} = W^\mathpzc{l}x_i^{\mathpzc{l} - 1} + b^\mathpzc{l}.
    \label{eq:Euclidean gcn feature transform}
\end{equation}
And, for neighbourhood aggregation, we would have:

\begin{equation}
    x_i^{\mathpzc{l}} = \sigma \left(e_i^\mathpzc{l} + \sum\limits_{j \in N(i)} w_{ij}e_i^\mathpzc{l} \right),
    \label{eq: euclidean aggregation}
\end{equation}
where, $N(i)$ denotes the number of neighbourhoods for node $i$. Note that $w_{ij}$ can be computed by a few different mechanisms  \cite{velivckovic2017graph, kipf2016semi}.
Therefore, to aggregate a neighbourhood in hyperbolic space, we have:
\begin{equation}
    A(h_i)= exp_{h_i}^\mathpzc{K} \left( \sum\limits_{j \in N(i)} w_{ij}log_{h_i}^\mathpzc{K}(h_j) \right).
\end{equation}
Hence, based on Eq. \ref{eq:Euclidean gcn feature transform}, Eq. \ref{eq:Weight matrix} and Eq. \ref{eq:bias translation}, the message passing in a hyperbolic GCN
can be expressed as follows:

For the hyperbolic feature transform in layer  $\mathpzc{l}$:
\begin{equation}
    h_i^{\mathpzc{l}} = (W \otimes x_i^{\mathpzc{l-1}}) \oplus b^{\mathpzc{l}},
\end{equation}
for the hyperbolic neighbourhood aggregation, we have:
\begin{equation}
    y_i^{\mathpzc{l}} = A(h^\mathpzc{l})_i.
\end{equation}

Thus, given a graph $\mathpzc{G} = (X,A)$ and its data augmentation $\mathpzc{G'} = (X', A')$, plus a hyperbolic GCN encoder  $\mathpzc{g}(\cdot)$ based on the above hyperbolic message, we have:
\begin{equation}
    \mathpzc{H} = \mathpzc{g}(\mathpzc{G}),
    \label{eq:hyperbolic GCN embedding}
\end{equation}
and
\begin{equation}
    \mathpzc{H'} = \mathpzc{g}(\mathpzc{G'}).
    \label{eq:hyperbolic GCN embedding_A}
\end{equation}

\subsubsection{Contrastive Loss Function} \label{ULoss}

\begin{algorithm}[!htbp]
\SetKwData{Left}{left}
\SetKwData{This}{this}
\SetKwData{Up}{up}
\SetKwFunction{Union}{Union}
\SetKwFunction{FindCompress}{FindCompress}
\SetKwInOut{Input}{Input}
\SetKwInOut{Output}{Output}
\caption{\textbf{UHSED}: Unsupervised Hyperbolic Social Event Detection}
\label{alg:UHSED}
\KwIn{A social message database $M=\{m_1,...,m_N\}$; A set of labels $\{l_0, l_1,...,l_n\}$; Hyperbolic MLP encoder $\mathpzc{g}(\cdot)$; Graph data augmentation method$AUG(\cdot)$; Readout function $\mathpzc{R}(\cdot)$; Contrastive loss function $\mathpzc{L}_{UHSED}(\cdot)$.}
\KwOut{A set of predicted social event labels $\{l_0,l_1',...,l_n'\}$, where $n$ is the number of classes.}
\BlankLine
\For{$m_i \in M$}
{Homogeneous message network $\mathpzc{G} = (X, A) \leftarrow$ data processing (Section \ref{DP})}
Negative sample $\mathpzc{G'} = AUG(\mathpzc{G})$\\
Hyperbolic embeddings for two graph: $\mathpzc{H} = \mathpzc{g}(\mathpzc{G})$ and $\mathpzc{H'} = \mathpzc{g}(\mathpzc{G'})$\\
Mapping hyperbolic embeddings into Euclidean space: $\mathpzc{E} = exp_o^\mathpzc{K} (\mathpzc{H})$ and $\mathpzc{E'} = exp_o^\mathpzc{K} (\mathpzc{H'})$\\
Readout:$\mathpzc{Z} = \mathpzc{R}(\mathpzc{E})$\\
\For{$e_i \in \mathpzc{E}$ and $e_i' \in \mathpzc{E'}$}
{ $\mathpzc{L}_{UHSED} \leftarrow \mathpzc{L}_{UHSED}(e_i, e_i', \mathpzc{Z}) $ \\
Update parameters}
$\{l_0',l_1',...,l_n'\} \leftarrow$ classify $\mathpzc{E}$
\end{algorithm}

When we have the graph representation of positive and negative samples. Before performing any downstream tasks, the hyperbolic embeddings need to be mapped to Euclidean space and a readout function  $\mathpzc{R}(\cdot)$ needs to be used to summarise the node embeddings into graph-level representations.

The hyperbolic embeddings can be mapped into Euclidean space via an  $exp$ mapping:
\begin{equation}
    \mathpzc{E} = exp_o^\mathpzc{K} (\mathpzc{H}),
\end{equation}
and
\begin{equation}
    \mathpzc{E'} = exp_o^\mathpzc{K} (\mathpzc{H'}).
\end{equation}

Here, the readout function  $\mathpzc{R}(\cdot)$ for embedding set $\mathpzc{E}$ is denoted as:
\begin{equation}
    \mathpzc{Z} = \mathpzc{R}(\mathpzc{E}),
\end{equation}
\begin{equation}
    \mathpzc{R}(\mathpzc{E}) = \sigma \left(\frac{1}{N} \sum\limits_{i=1}^N e_i \right),
\end{equation}
where $e_i \in \mathpzc{E}$.

Then, a discriminator  $\mathpzc{D}(\cdot)$ \cite{oord2018representation} is used to maximise the mutual information, denoted as:
\begin{equation}
    \mathpzc{D}(e_i,\mathpzc{Z}) = \rho \left(e_i \mathpzc{W} \mathpzc{Z} \right),
\end{equation}
where $\mathpzc{W}$ denotes a learnable scoring matrix, and $\rho$ is nonlinear logistic sigmoid.

A contrastive loss function  $\mathpzc{L}_{UHSED}(\cdot)$ is defined to maximise the mutual information between  $\mathpzc{Z}$ and $\mathpzc{Z'}$. Here we employ BCE loss, and so we have:
\begin{equation}
\footnotesize
    \mathpzc{L}_{UHSED} = \frac{1}{N+M} \left(
    \sum\limits_{i=1}^N \mathpzc{E}[log\mathpzc{D}(e_i,\mathpzc{Z}) + 
    \sum\limits_{j=1}^M \mathpzc{E'}[log (1-\mathpzc{D}(e_j,\mathpzc{Z}))]
    \right).
\end{equation}

Lastly, to detect the social events, a simple linear classifier (logistic regression) is used to classify the unsupervised training results $\mathpzc{E}$. The pseudocode for the UHSED model is given in Algorithm  \ref{alg:UHSED}.

\section{Experiments}
The experiments were designed to answer the following questions:
\begin{itemize}
\item $Q1$: Do the proposed approaches have superiority over the baseline models? (Sections \ref{qs:HSED_baseline}, \ref{qs:UHSED_baselines})

\item $Q2$: What is the impact of the hidden layers and their dimensionality on the proposed models? (Sections \ref{qs:HSED_parameters}, \ref{qs:UHSED_parameters})

\item $Q3$:Which data augmentation method is the most suitable for the UHSED model? (Section \ref{qs:UHSED_data_augmentation})

\item $Q4$: Are hyperbolic spaces better for social media data than Euclidean spaces? (Section \ref{qs:hyperbolic spaces})

\item $Q5$: Which hyperbolic model performs better with social media data? (Section \ref{qs:hyperbolic model})
\item $Q6$: Does the tree-like structure of the Twitter dataset hinder neighbour aggregation? (Section \ref{qs: tree-like GCN})

\end{itemize}

\subsection{Datasets}
Our model is mainly aimed at social media data, so we chose the Twitter \cite{mcminn2013building} real-world public dataset as our primary dataset. However, since the Twitter dataset is a large-scale dataset, performing graph contrastive learning with it might lead to out-of-memory errors. Thus, we created a balanced mini-Twitter dataset based on the larger original dataset. Additionally, we experimented with the Cora \cite{sen2008collective} dataset, the Citeseer \cite{sen2008collective} dataset and the UHSED model to evaluate the efficacy of modelling in hyperbolic space. An overview of the datasets used is shown in Table \ref{tab:datasets}.

\begin{table}[htbp]
\footnotesize
\caption{The statistical information of datasets.}\label{tab:datasets}
\centering
\vspace{0mm}
\setlength{\tabcolsep}{1mm}{
\scalebox{1}{
\begin{tabular}{cccccc}
\multicolumn{6}{c}{\textit{\textbf{Dataset for HSED}}}                                                                                              \\ \hline
Dataset                                                                       & Num. of Classes & \multicolumn{2}{c}{Num. of Nodes} &  Num. of Features \\ \hline
Twitter                                                                       & 503             & \multicolumn{2}{c}{68,841}        &  302  \\ \hline
                                                                                     &         &                   &                &          \\
\multicolumn{6}{c}{\textit{\textbf{Datasets for UHSED}}}                                                                                               \\ \hline
Dataset                                                                      & Num. of Classes & \multicolumn{2}{c}{Num. of Nodes} &  Num. of Features \\ \hline
mini-Twitter                                                                 & 15              & \multicolumn{2}{c}{3,000}         & 302    \\ 
Cora                                                                         & 7               & \multicolumn{2}{c}{2,708}         & 1,433    \\
Citeseer                                                                     & 6               & \multicolumn{2}{c}{3,327}         & 3,703    \\ \hline
                                                                        &             &         &                   &               &          \\

\end{tabular}
}
}
\end{table}

\subsection{Experiments on HSED in Supervised Scenarios}
This section presents the evaluations of our HSED model. Section  \ref{HSED_baselines} introduces the baseline models, and Section \ref{HSED_setting} outlines the experimental environment and the hyperparameter settings. The rest of this section provides the details of the experiments and answers research questions $Q1$ and $Q2$.
\subsubsection{Baselines} \label{HSED_baselines}
To evaluate the HSED model, we conducted comprehensive experiments with the Twitter dataset and compared the results to other models – some traditional and some
start-of-art.  The source code of the proposed HSED model can be found on Github\footnote{https://github.com/ZITAIQIU/HSED}. The compared baselines include: 
\begin{itemize}
    \item \textbf{Word2vec} 
\cite{mikolov2013efficient} – a message representation learning model widely used in many social event detection models.
    \item \textbf{LDA} \cite{blei2003latent} – a traditional topic detection model. Its statistical approach captures topic-related co-occurrence words in messages.
    
    \item \textbf{WMD} \cite{kusner2015word} – a similarity measurement method that detects social events by calculating the similarity between messages.
    
    \item \textbf{BERT} \cite{devlin2018bert} – a powerful language representation model that plays a key role in many start-of-the-art social event detection models.

    \item \textbf{KPGNN} \cite{cao2021knowledge} – a social event detection model based on HINs. It mainly focuses on incremental social event detection, and its offline performance is slightly better than PP-GCN’s.

    \item \textbf{FinEvent} \cite{peng2022reinforced} – a social event detection model based on incremental and cross-lingual social messages. It mainly applies reinforcement learning to improve the performance of social event detection.
\end{itemize}

\subsubsection{Parameter Settings} \label{HSED_setting}
To ensure reliable results, we ran each experiment five times, reporting the average results as final. The parameter settings for HSED are shown in Table  \ref{tab:HSED_parameters}.

\textit{Hidden layer} denotes the number of hidden layers. \textit{Hidden dimension} stands for the dimensions of the hidden layers. Note that the  \textit{training, test and validation rate}  were designed based on the offline experimental settings for KPGNN\footnote{https://github.com/RingBDStack/KPGNN}. All the experiments with the HSED model were conducted on an NVIDIA V100 GPU with 12 24-core Intel Xeon Scalable ‘Cascade Lake’ processors.

\begin{table}[!htb]
    \centering
    \footnotesize
    \caption[HSED parameter setting]{HSED parameter settings.}
    \scalebox{1.2}{
    \begin{tabular}{cc}
    \toprule[1 pt]
    Parameter & Value  \\ \midrule
    \textit{Hidden layer} & 2  \\
    \textit{Hidden dimension} & 512   \\ 
    \textit{Training rate} & $70\%$   \\ 
    \textit{Test rate} & $20\%$   \\ 
    \textit{Validation rate} & $10\%$   \\ 
    \textit{Learning rate} & 0.1  \\
    \textit{Optimiser} & Adam   \\
    \textit{Activation function} & ReLU   \\
  
    \bottomrule[1pt]
    \end{tabular}}
    \label{tab:HSED_parameters}
\end{table}

\subsubsection{Evaluation Metrics}
In some baseline models that use clustering algorithms to
cluster social events (like KPGNN), the labels predicted by the models will differ from the ground-truth label. Hence, standard accuracy metrics are not a suitable evaluation measure. Instead, we used the same evaluation metrics as the baseline models to measure the similarity between the predicted and the ground-truth
labels to ensure a fair comparison. Namely, these were NMI  \cite{estevez2009normalized}, AMI \cite{xuan2010information}, and ARI \cite{xuan2010information}, which have been widely used to assess social event detection models \cite{cao2021knowledge,peng2019fine}.

\subsubsection{HSED Model Performance Comparison (Answer Q1)} \label{qs:HSED_baseline}

The experimental results of the HSED model and the baseline models with the Twitter dataset are shown in Table  \ref{tab:HSED result}. Overall, the HSED model outperformed the baseline models on every metric. We also found a high degree of agreement between the labels predicted by HSED and the ground-truth labels. However, none of the baseline models could capture enough information from the social networks to compete with HSED. For example, LDA only measures topic-related co-occurrence words in messages, while Word2Vec, WMD, and BERT only consider message embeddings. These models ignore the rich semantic and structural information in heterogeneous networks. Although KPGNN and FinEvent learn the heterogeneity of social media based on previous models, they ignore the fact that social media data is tree-structured. Particularly with such a large-scale dataset, they were not able to discern the differences between the leaves, as shown in Fig. \ref{fg:HE_distance}. Thus, the baseline models only reached $80\%$ NMI, $69\%$ AMI and $48\%$ ARI at the highest. By contrast, HSED considers both heterogeneity and the exponential growth of social media data. In view of this, the model was able to learn more information than the baseline models and so scored highest in all the metrics.

\begin{table}[!htb]
    \centering
    \caption[HSED classification results]{Comparison experiment results of classification of all models. }
    \scalebox{0.8}{
    \begin{tabular}{c|cccccc|c}
    \toprule[1 pt]
    Metrics & Word2Vec & LDA & WMD & BERT  &KPGNN & FinEvent & HSED (ours) \\ \midrule
    NMI & 0.44 & 0.29 & 0.65 & 0.64 & 0.70  & 0.80 & \textbf{0.96} \\
    AMI & 0.13 & 0.04 & 0.50 & 0.44 & 0.52 & 0.69 & \textbf{0.83} \\ 
    ARI & 0.02 & 0.01 & 0.06 & 0.07 & 0.22  & 0.48 & \textbf{0.94} \\
  
    \bottomrule[1pt]
    \end{tabular}
    }
    \label{tab:HSED result}
\end{table}

\subsubsection{The Impact of The Hidden Layers and Their Dimensionality on The HSED Model (Answer Q2)} \label{qs:HSED_parameters}

\begin{figure*}[!tb]
\centering
\subfigure[ACC]{\includegraphics[width=0.19\textwidth]{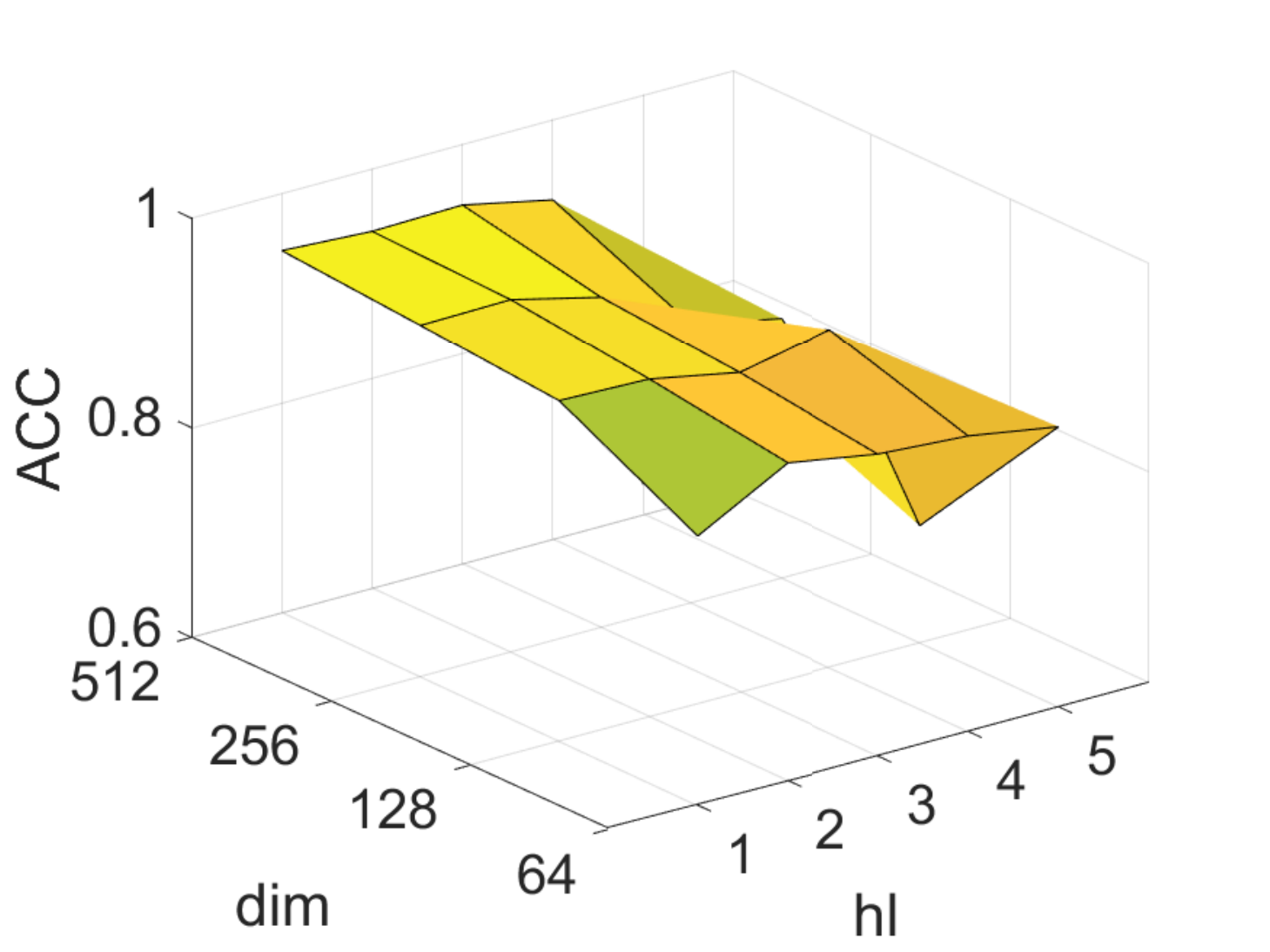}}
\subfigure[NMI]{\includegraphics[width=0.19\textwidth]{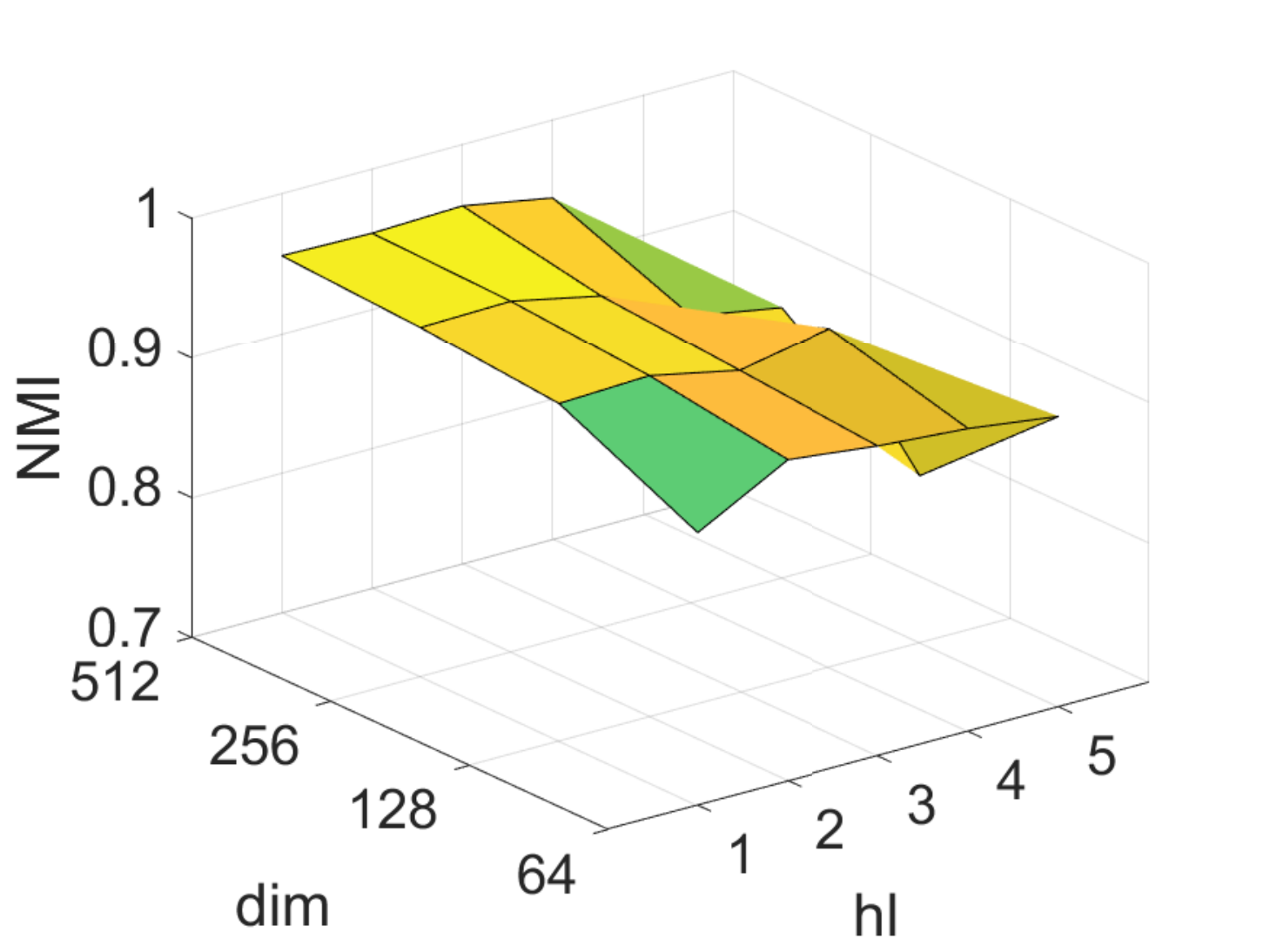}}
\subfigure[AMI]{\includegraphics[width=0.19\textwidth]{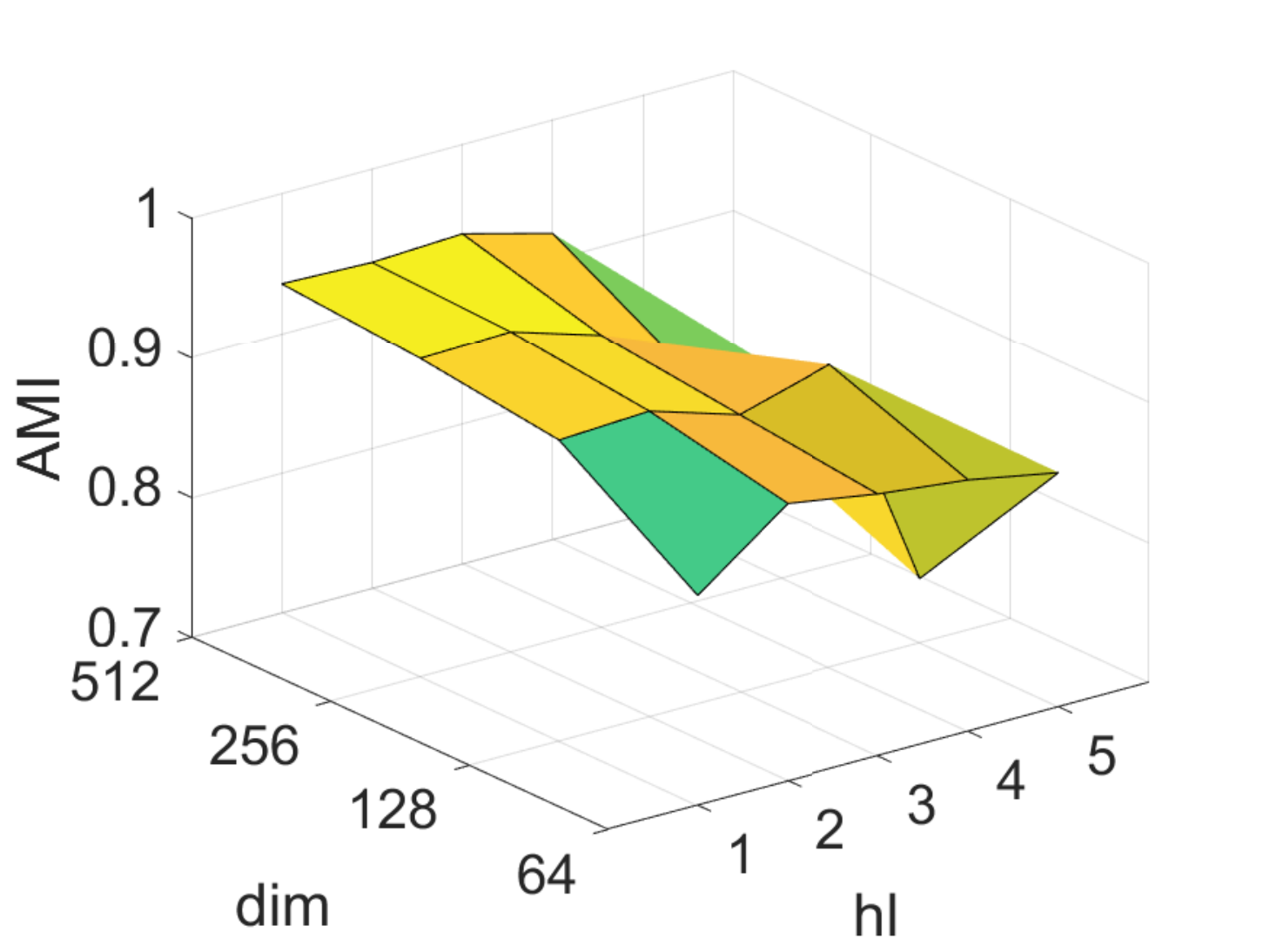}}
\subfigure[ARI]{\includegraphics[width=0.19\textwidth]{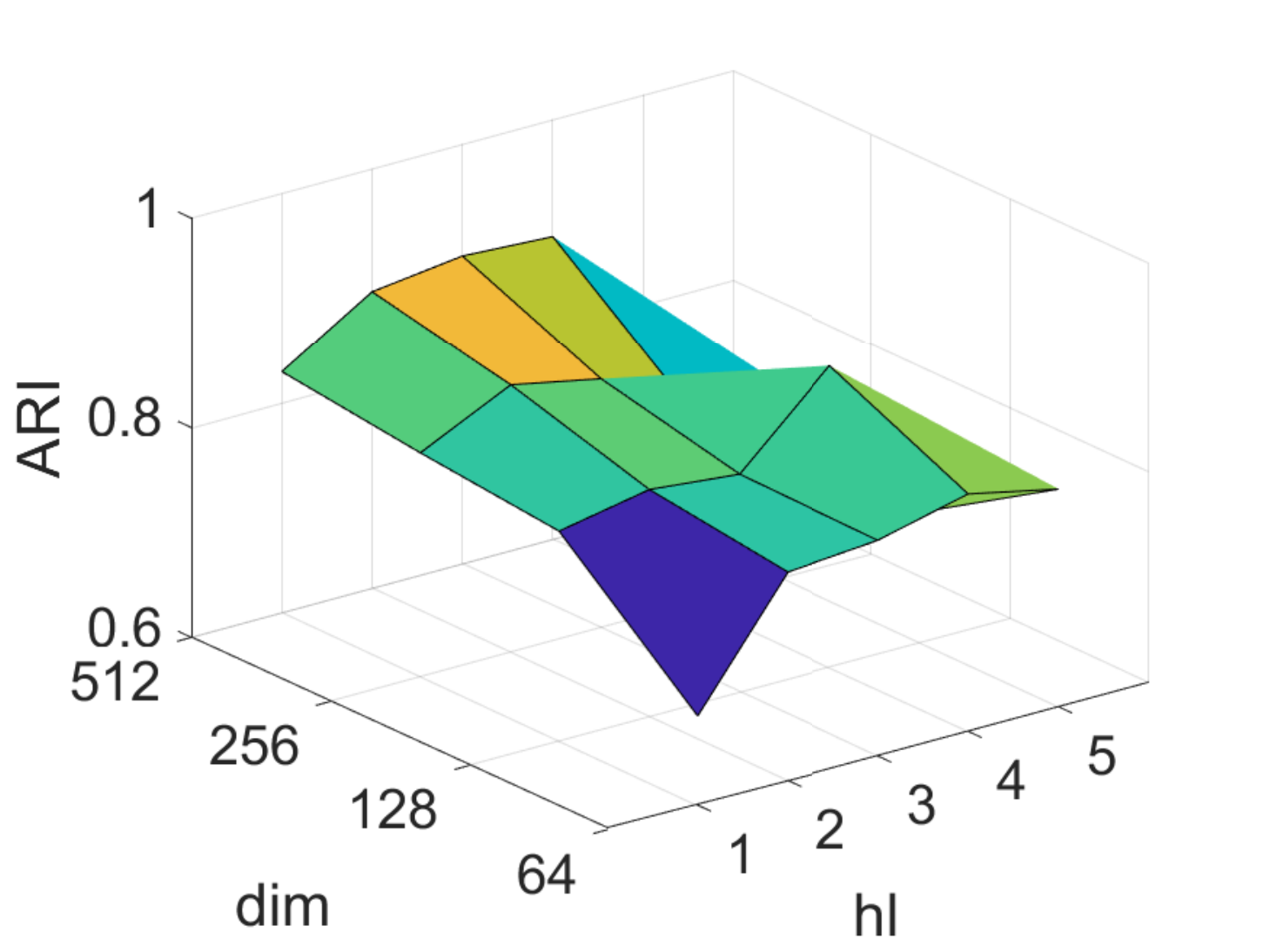}}
\subfigure[Time]{\includegraphics[width=0.19\textwidth]{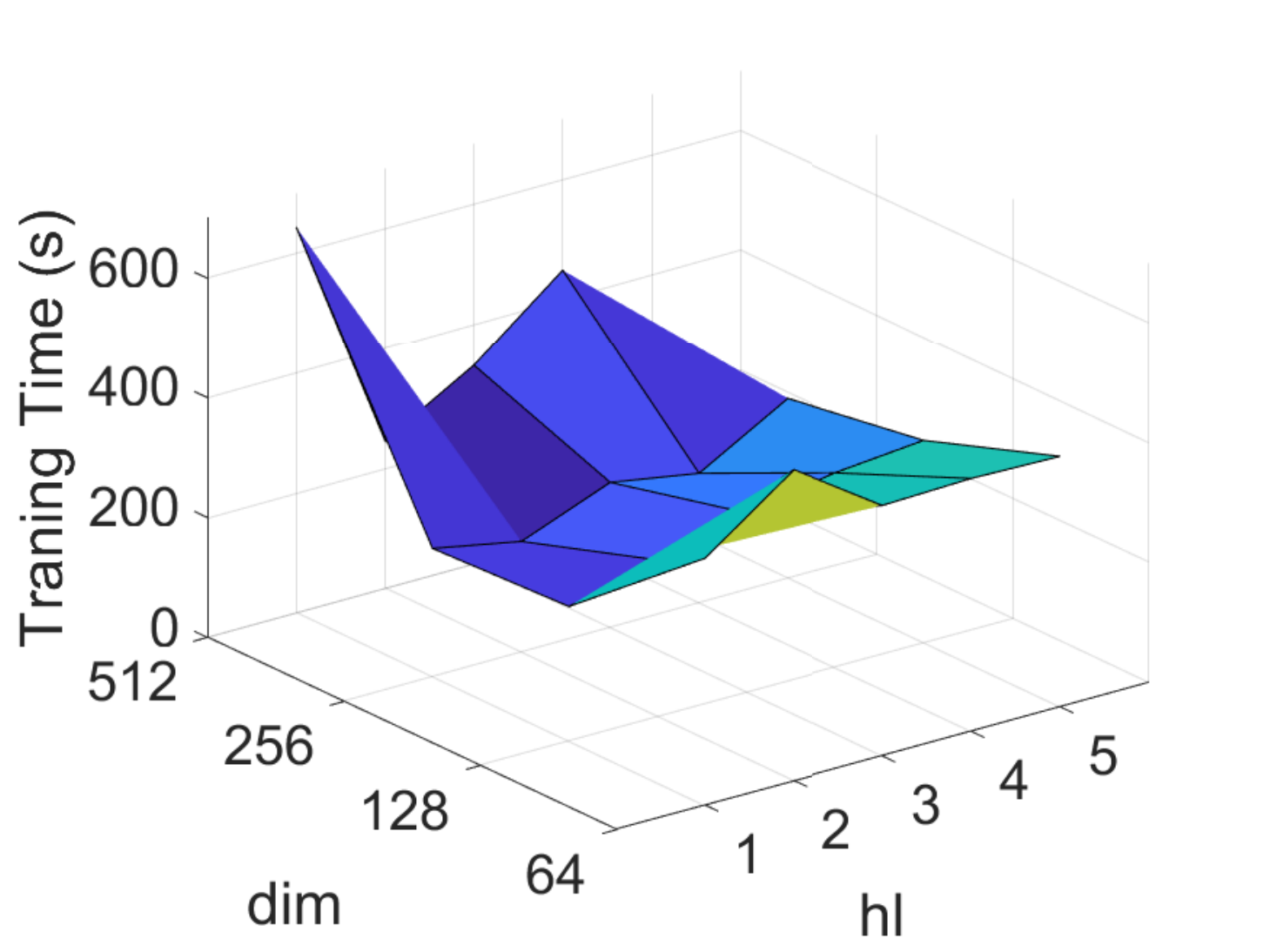}}

\caption[HSED with different hyperparameters]{HSED with different hyperparameters. “dim” is the hidden dimension and “hl” is the hidden layer.}

\label{fg:HSED_para_analysis}
\end{figure*}

The HSED model is based on an MLP. Although the HSED model partly operates in hyperbolic space, it still has the number of hidden layers and the dimensionality of those layers as its hyperparameters. In this section, we study the impact of these two hyperparameters on the HSED model. We designed experiments to test the effects of these hyperparameters – the results of which are shown in Fig.  \ref{fg:HSED_para_analysis}.

As the figure shows, the higher the number of dimensions at, say 256 or 512, the better the performance. With too few dimensions, i.e., 64, performance degraded. However, the dimensionality of the hidden layers is not the most critical factor affecting HSED’s performance – the number of hidden layers is. With more than three hidden layers, HSED’s performance degraded rapidly across all metrics. 
In addition, we also explored the effect of changes in these two parameters on the model’s training time. The results, appearing in Fig. \ref{fg:HSED_para_analysis}(e), shows that neither parameter has much influence over training time. Therefore, for the best performance, one should include less than three hidden layers with each layer having a relatively high dimensionality.

\subsection{Experiments on UHSED in Unsupervised Scenarios}
In this section, we evaluate the UHSED model. Section  \ref{USED_baselines} introduces the baselines. Section  \ref{UHSED_setting} provides details of the experimental environment and hyperparameter settings. The rest of this section outlines the experiments and answers research questions $Q1$, $Q2$, and $Q3$ for the UHSED model.
\subsubsection{Baselines} \label{USED_baselines}
To evaluate our UHSED model, we conduct comprehensive experiments with the mini-Twitter, Cora, and Citeseer datasets and compared the results with the current state-of-the-art models. The baselines included:
\begin{itemize}
    \item \textbf{DGI} \cite{velivckovic2017graph} – a single-branch graph contrastive learning model. Negative samples are obtained through data augmentation or corrupting the original image.
    
    \item \textbf{GraphCL} \cite{you2020graph} – a dual-branch graph contrastive learning model that augments the original graph twice to obtain two views.
    
\end{itemize}

\subsubsection{Parameter Settings} \label{UHSED_setting}

The parameter settings for UHSED are shown in Table  \ref{tab:UHSED_parameters}.

\textit{Hidden layer} stands for the number of hidden layers in the UHSED model.  \textit{Hidden dimension} is the dimensionality of the hidden layers.  \textit{Drop rate} is the percentage of features or nodes that are dropped through data augmentation. \textit{Augmentation type} is the method of graph data augmentation (see Table \ref{tab:graph_data_a} for more details). All the experiments with the UHSED model were conducted on an NVIDIA V100 GPU with 12 24-core Intel Xeon Scalable ‘Cascade Lake’ processors.

\begin{table}[!htb]
    \centering
    \footnotesize
    \caption[UHSED parameters settings]{UHSED parameter settings.}
    \scalebox{1.2}{
    \begin{tabular}{cc}
    \toprule[1 pt]
    Parameter & Value  \\ \midrule
    \textit{Hidden layer} & 1  \\
    \textit{Hidden dimension} & 512   \\ 
    \textit{Drop rate} & $10\%$   \\ 
    \textit{Learning rate} & 0.1  \\
    \textit{Optimiser} & Adam   \\
    \textit{Activation function} & ReLU   \\
    \textit{Augmentation method} & Feature corruption   \\
  
    \bottomrule[1pt]
    \end{tabular}
    }
    \label{tab:UHSED_parameters}
\end{table}
\subsubsection{Evaluation Metrics} 
Since UHSED is an unsupervised learning model, we employed Micro-F1 and Macro-F1 to evaluate the accuracy of the detection results following previous studies  \cite{velivckovic2017graph,you2020graph}.


\subsubsection{UHSED Model Performance Comparison (Answer Q1)} \label{qs:UHSED_baselines}

The experimental results for the UHSED model and baselines with the mini-Twitter,
Cora and Citeseer datasets are shown in Table \ref{tab:UHSED results}. Overall, the UHSED model outperformed the baseline models. Initially, our experiments only included the mini-Twitter dataset. However, we noticed that all models returned relatively low scores. So, to rule out problems with the model’s code, we added the Cora and Citeseer datasets for validation. On these two datasets, the models yielded typical performance. Therefore, we wondered whether the low scores were due to the tree-like structure of the Twitter dataset and whether perhaps this was hindering the neighbour aggregation process in the GCN. To this end, we undertook further experiments, discussing the results in Section \ref{qs: tree-like GCN}. 

In addition, we noted that the DGI model outperformed the GraphCL model. Given that UHSED also has a single-branch architecture, this may indicate that a single-branch model is more suitable for graph contrastive learning. Regardless, the results make it clear that hyperbolic spaces do help the model to capture more information from tree-structured data. Section \ref{qs:hyperbolic spaces} provides a lengthier discussion on this phenomenon.

\begin{table}[!t]
\centering
\small
\renewcommand\arraystretch{1}
\setlength{\tabcolsep}{1.5mm}
\caption[UHSED classification results]{Comparison experiment results of classification of UHSED and baseline models.}
\scalebox{0.8}{
\begin{tabular}{ccccccc} 
\toprule[1pt]
\multirow{2}{*}{Methods} & \multicolumn{2}{c}{mini-Twitter} & \multicolumn{2}{c}{Cora} & \multicolumn{2}{c}{Citeseer} \\ 
\cmidrule(lr){2-3}\cmidrule(lr){4-5}\cmidrule(lr){6-7}
 & Micro-F1 & Macro-F1 & Micro-F1 & Macro-F1 & Micro-F1 & Macro-F1 \\ \midrule
DGI & $\underline{0.1714}$ & $\underline{0.14129}$ & $\underline{0.8077}$ & $\underline{0.7922}$ & $\underline{0.6885}$ & $\mathbf{0.6518}$ \\
GraghCL & $0.1517$ & $0.1302$ & $0.6993$ & $0.6837$ & $0.6377$ & $0.5971$  \\
UHSED (ours) & $\mathbf{0.5288}$ & $\mathbf{0.5266}$ & $\mathbf{0.8314}$ & $\mathbf{0.8203}$ & $\mathbf{0.7081}$ & $\underline{0.6458}$  \\ 

 \bottomrule[1pt]
\end{tabular}}
\label{tab:UHSED results}
\end{table}

\subsubsection{The Impact of The Hidden Layers and Their Dimensionality on The UHSED Model (Answer Q2)} \label{qs:UHSED_parameters}

\begin{figure*}[!tb]
\centering
\subfigure[Micro-F1 on mini-Twitter]{\includegraphics[width=0.28\textwidth]{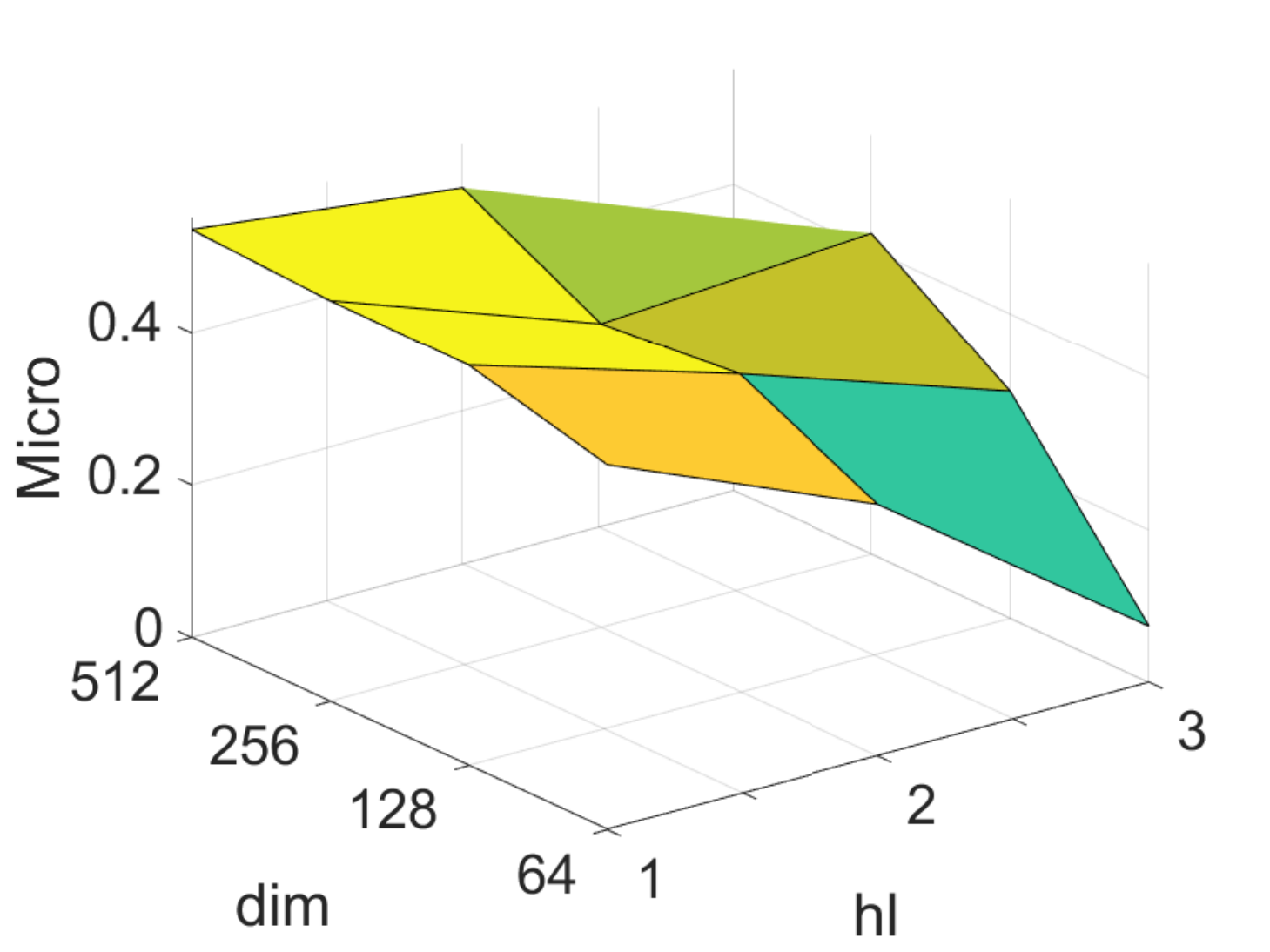}}
\subfigure[Macro-F1 on mini-Twitter]{\includegraphics[width=0.28\textwidth]{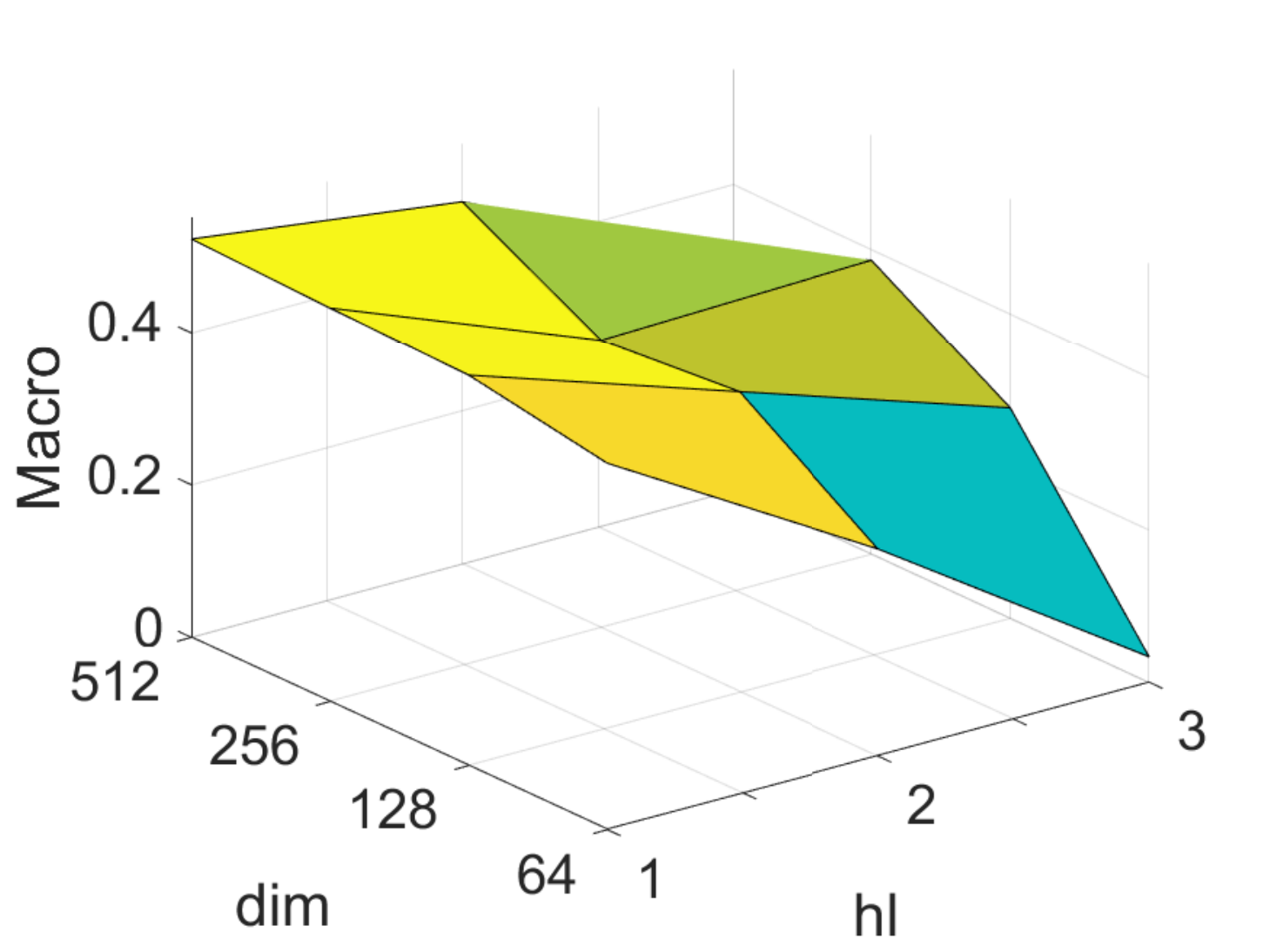}}
\subfigure[Time on mini-Twitter]{\includegraphics[width=0.28\textwidth]{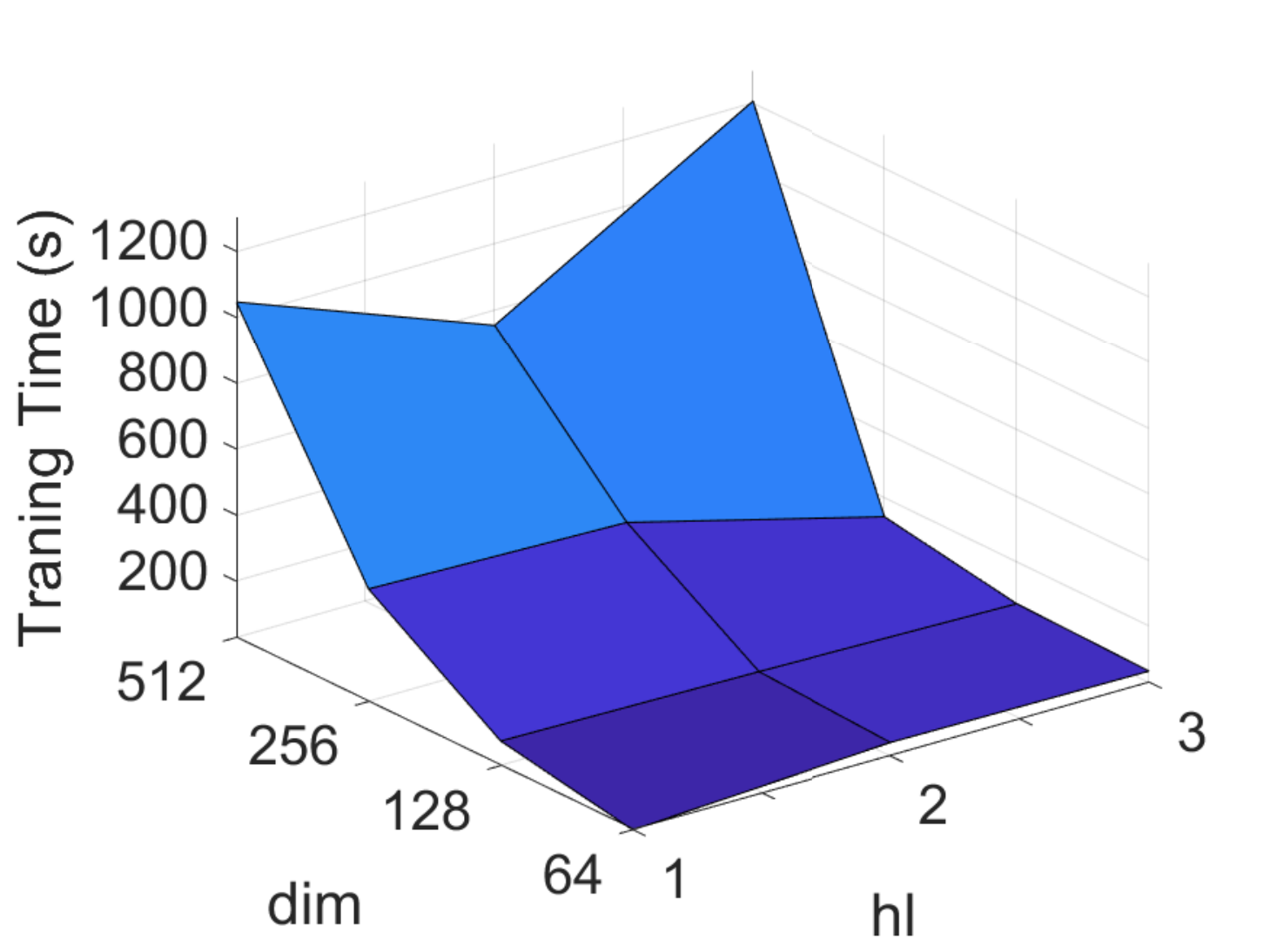}}

\subfigure[Micro-F1 on Cora]{\includegraphics[width=0.28\textwidth]{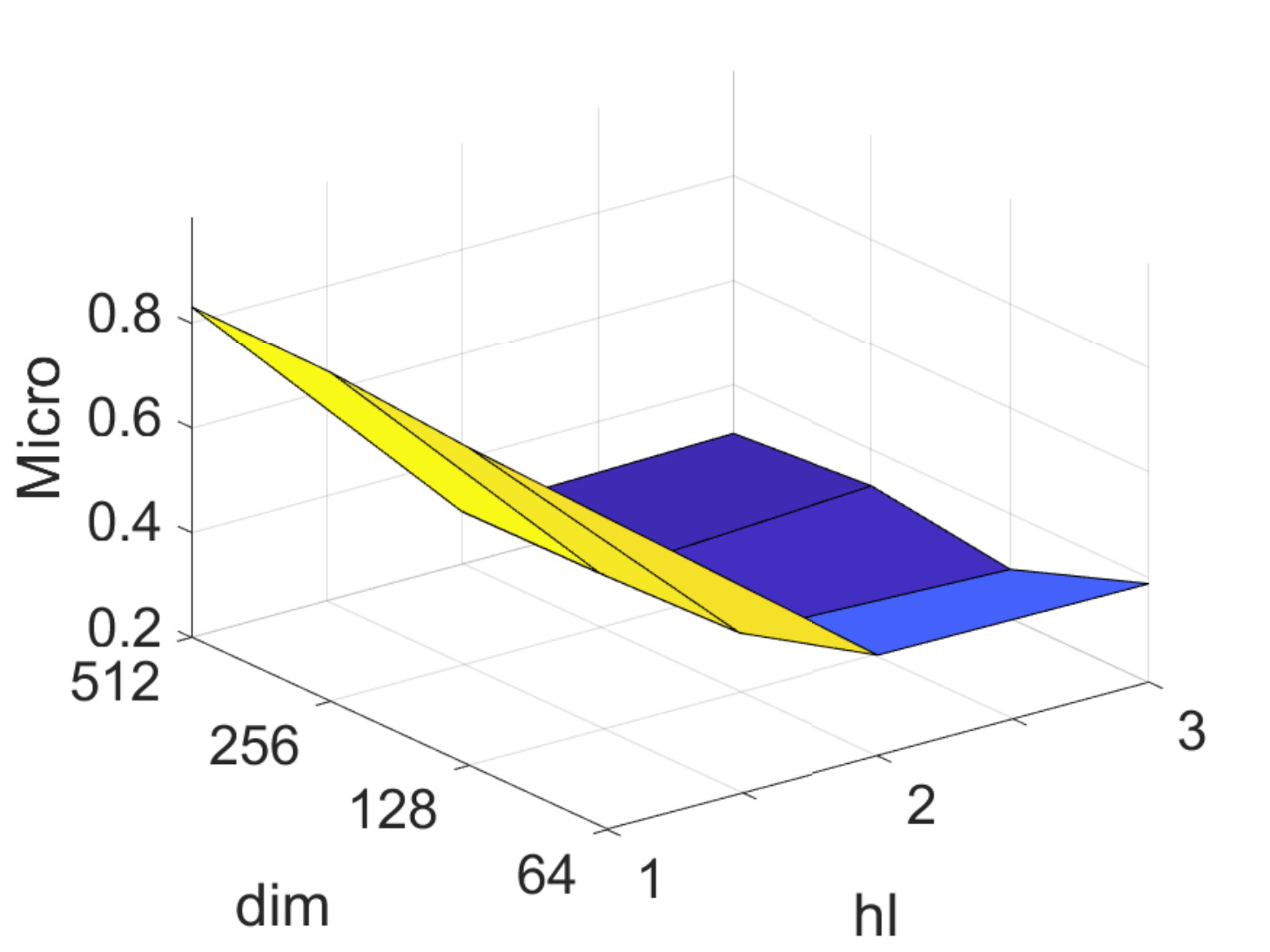}}
\subfigure[Macro-F1 on Cora]{\includegraphics[width=0.28\textwidth]{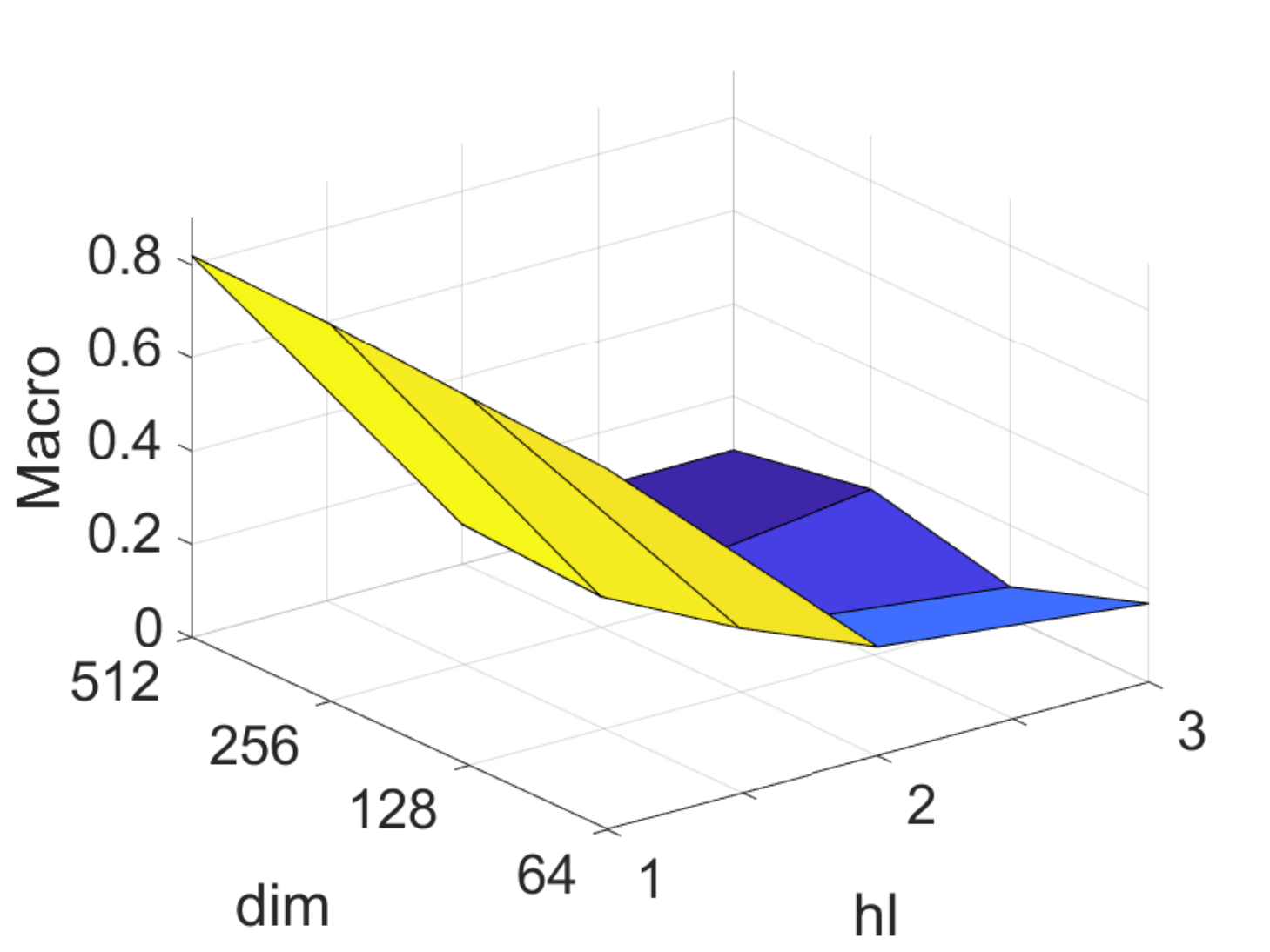}}
\subfigure[Time on Cora]{\includegraphics[width=0.28\textwidth]{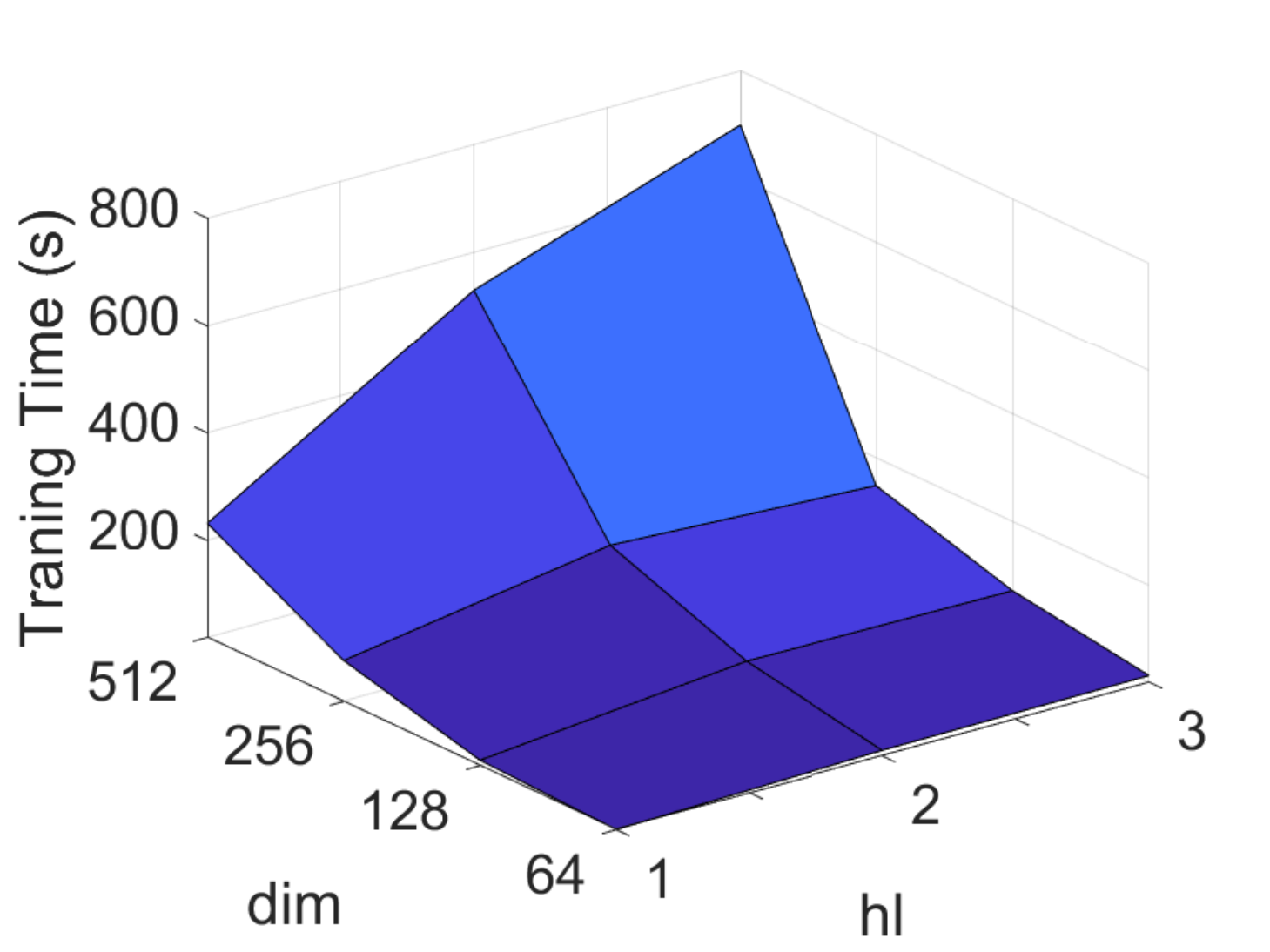}}

\subfigure[Micro-F1 on Citeseer]{\includegraphics[width=0.28\textwidth]{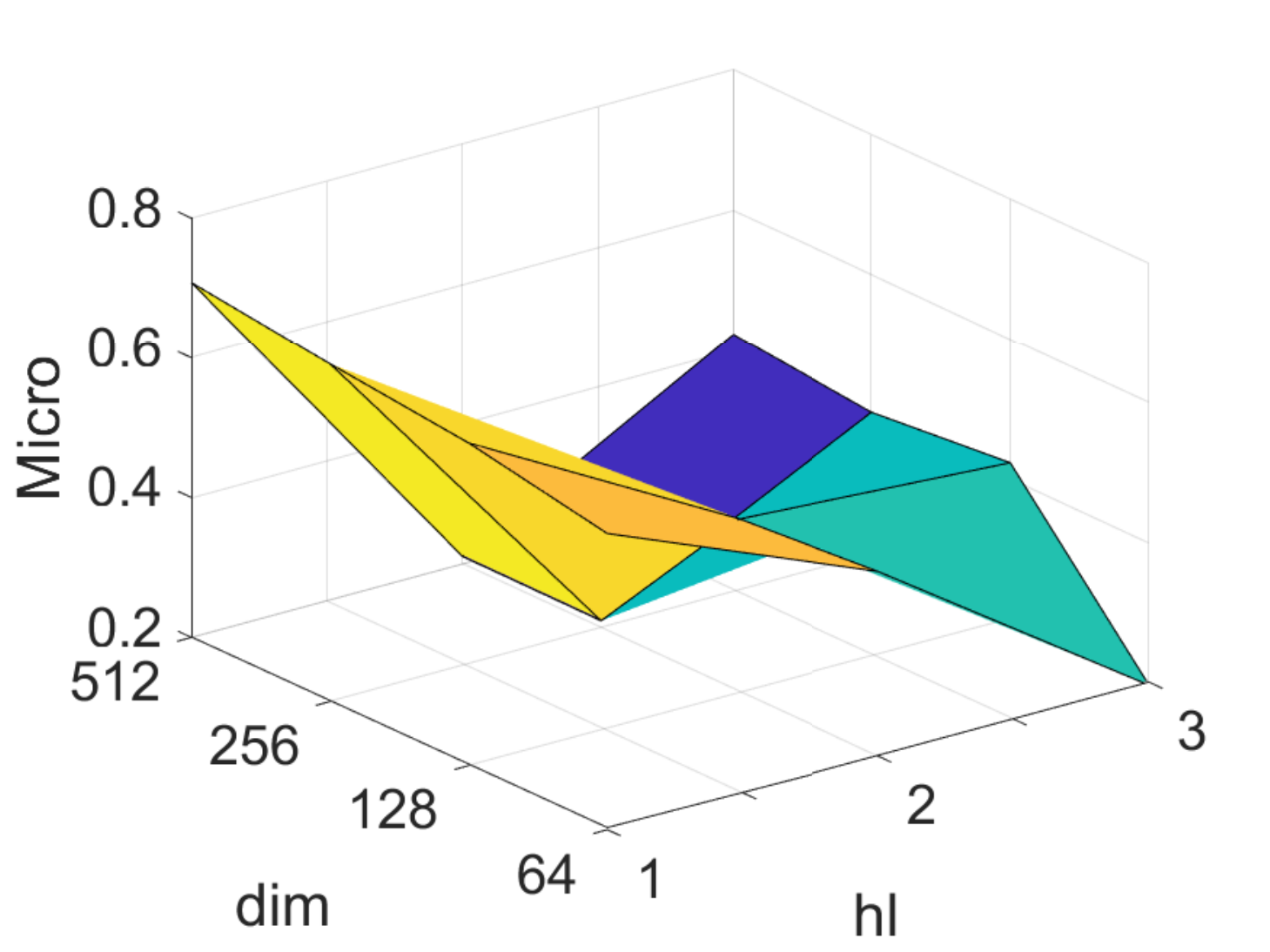}}
\subfigure[Macro-F1 on Citeseer]{\includegraphics[width=0.28\textwidth]{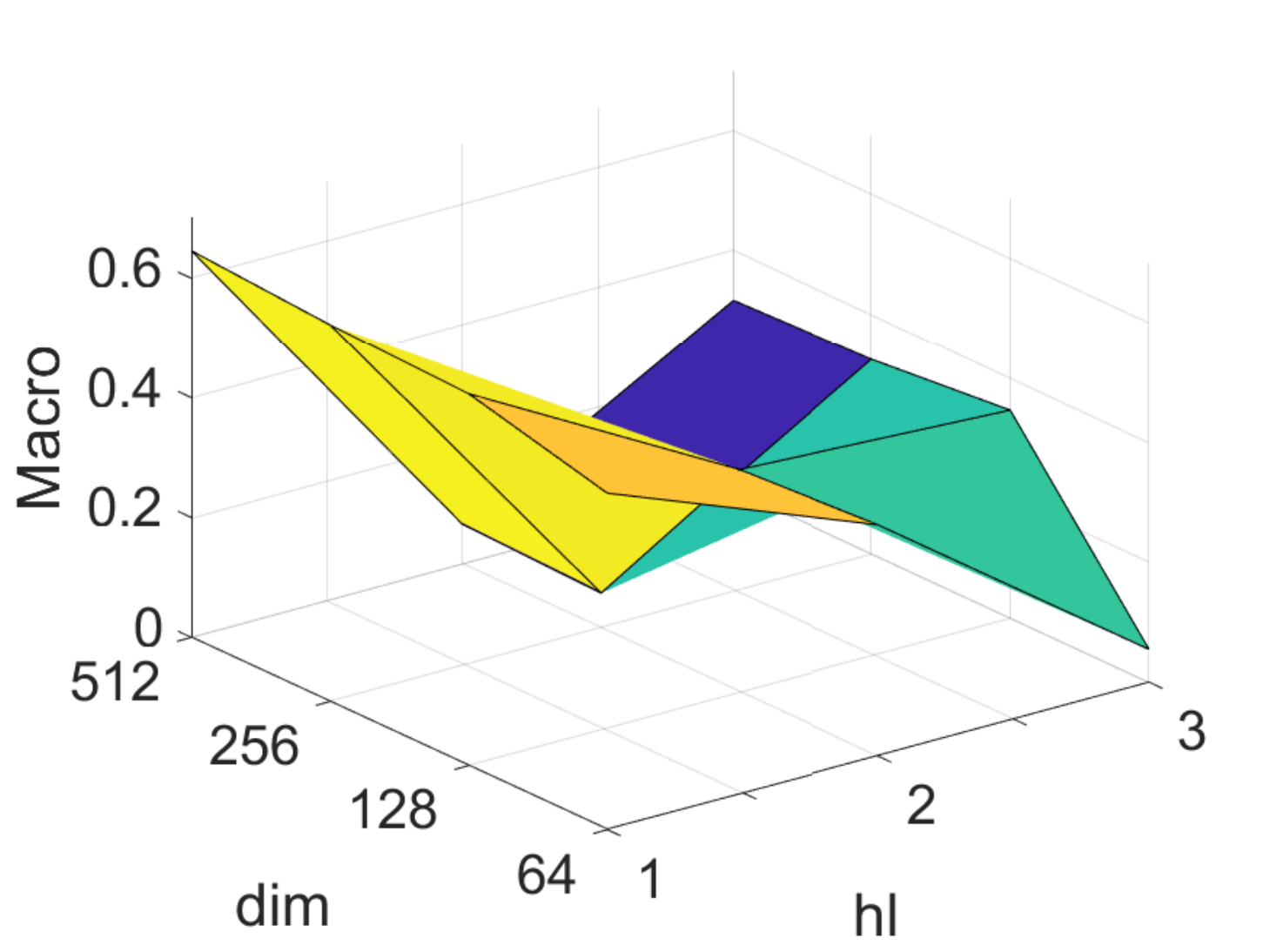}}
\subfigure[Time on Citeseer]{\includegraphics[width=0.28\textwidth]{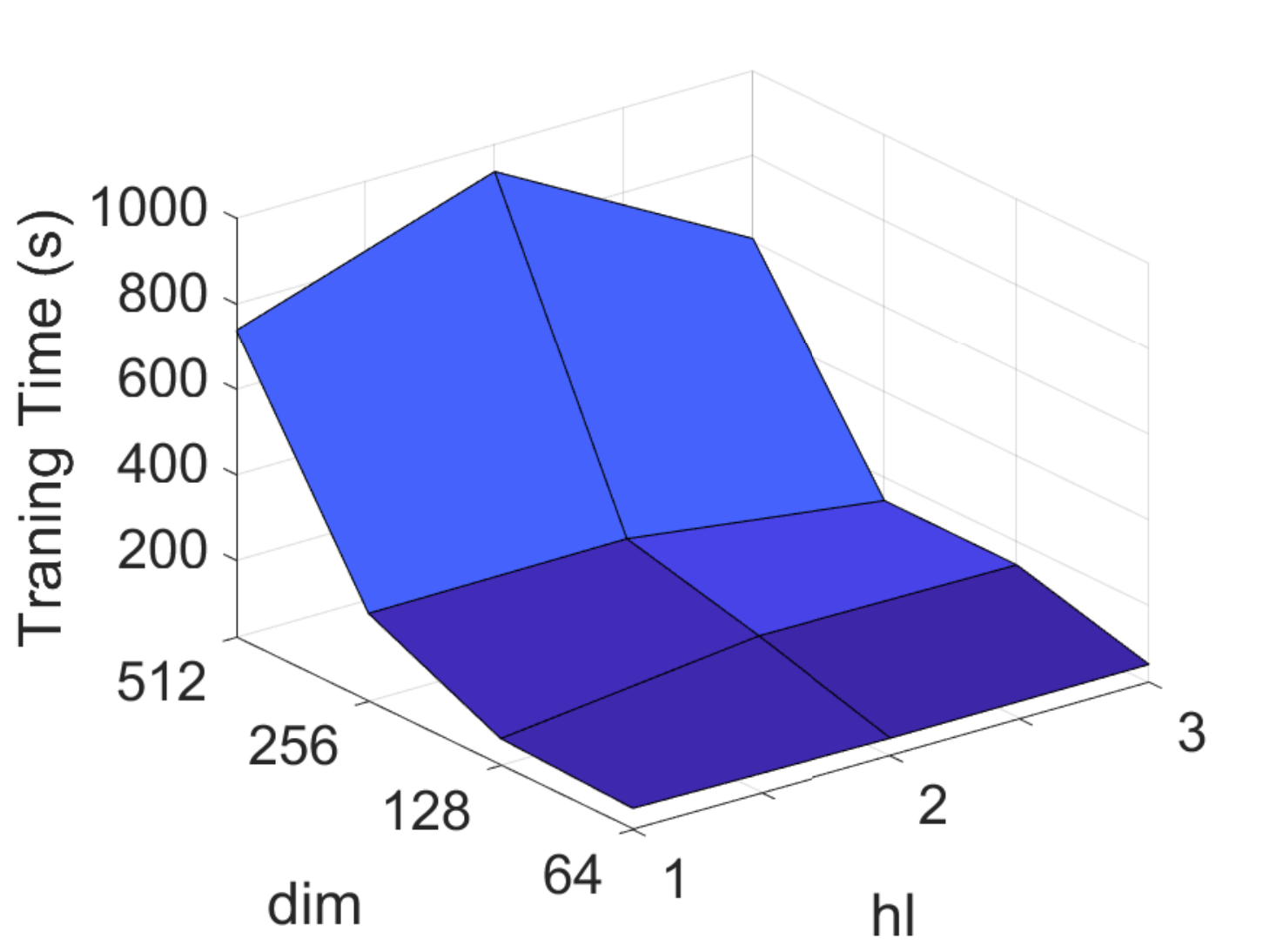}}

\caption[UHSED with different hyperparameters]{UHSED model with different hyperparameters.}
\label{fg:UHSED_para_analysis}
\end{figure*}

The UHSED model was developed from the HSED model. Thus, the same hyperparameters apply – being the number of hidden layers and their dimensionality. In this subsection, we explore the impact of these two hyperparameters on the UHSED model. The results are shown
in Fig.  \ref{fg:UHSED_para_analysis}.

Like the HSED model, we note that the higher the number of dimensions, the better the performance. However, the number of hidden layers had a great impact on UHSED’s performance with all datasets. From Fig. \ref{fg:UHSED_para_analysis} (a), (b), (d), (e), (g) and (h), UHSED’s scores drop rapidly with more than two hidden layers. The results for model training times are shown in Fig. \ref{fg:UHSED_para_analysis} (c), (f) and (i). Obviously, the number of hidden layers has little effect on the training time, but with a dimensionality of larger than 256, training time proliferated for all datasets. Thus, we suggest an architecture with only one hidden layer for all datasets and no more than 256 dimensions for large-scale datasets.

\subsubsection{Selection of data augmentation method in the UHSED Model (Answer Q3)} \label{qs:UHSED_data_augmentation}

The key idea behind contrastive learning is to generate negative samples through data augmentation. The UHSED model is compatible with three graph data augmentation methods. Hence, in this section, we explore these different data augmentation methods, comparing them for their effectiveness.

For these experiments, the drop rate of augmentation was set to  $0.1$, and the number of hidden dimensions was set to $512$ with one hidden layer. The results are shown in Table  \ref{tab:data_augmentation_analysis}.

From the results, we can see that  \textit{“Feature corruption”} outperformed the other graph data augmentation methods. The reason is that this method only disrupts the order of the node features; it does not drop any features, whereas the other two data augmentation methods do drop some information. The \textit{“Feature dropping”} method drops more feature information than the \textit{“Random masking”} method, which resulted in the \textit{“Feature dropping”} method getting lower scores than \textit{“Random masking”}. Furthermore, UHSED is a single-branch graph contrastive learning model, and we believe that data augmentation methods that drop features may not be suitable for a contrastive learning model with a single-branch architecture.

\begin{table}[t]
\centering
\small
\renewcommand\arraystretch{1}
\setlength{\tabcolsep}{1.5mm}
\caption[Graph data augmentation analysis]{ Graph data augmentation analysis.}
\scalebox{0.75}{
\begin{tabular}{ccccccc} 
\toprule[1pt]
\multirow{2}{*}{Methods} & \multicolumn{2}{c}{mini-Twitter} & \multicolumn{2}{c}{Cora} & \multicolumn{2}{c}{Citeseer} \\ 
\cmidrule(lr){2-3}\cmidrule(lr){4-5}\cmidrule(lr){6-7}
 & Micro-F1 & Macro-F1 & Micro-F1 & Macro-F1 & Micro-F1 & Macro-F1 \\ \midrule
Feature dropping & 0.4215 & 0.3812 & 0.3249 & 0.1222 & 0.3976 & 0.3434 \\
Random masking & 0.4744 & 0.4668 & 0.7588 & 0.7489 & 0.6123 & 0.5747  \\
Feature corruption & $\mathbf{0.5288}$ & $\mathbf{0.5266}$  & $\mathbf{0.8314}$ & $\mathbf{0.8203}$ & $\mathbf{0.7081}$ & $\mathbf{0.6458}$ \\ 

 \bottomrule[1pt]
\end{tabular}
}
\label{tab:data_augmentation_analysis}
\end{table}
\subsection{Discussion} \label{discussion}
In this section, we answer our remaining research questions.  
\subsubsection{Comparison of Model Performance in Hyperbolic Space and Euclidean Space (Answer Q4)} \label{qs:hyperbolic spaces}

\begin{figure}[!tb]
\centering
\subfigure[Hyperbolic space analysis]{\includegraphics[width=0.24\textwidth]{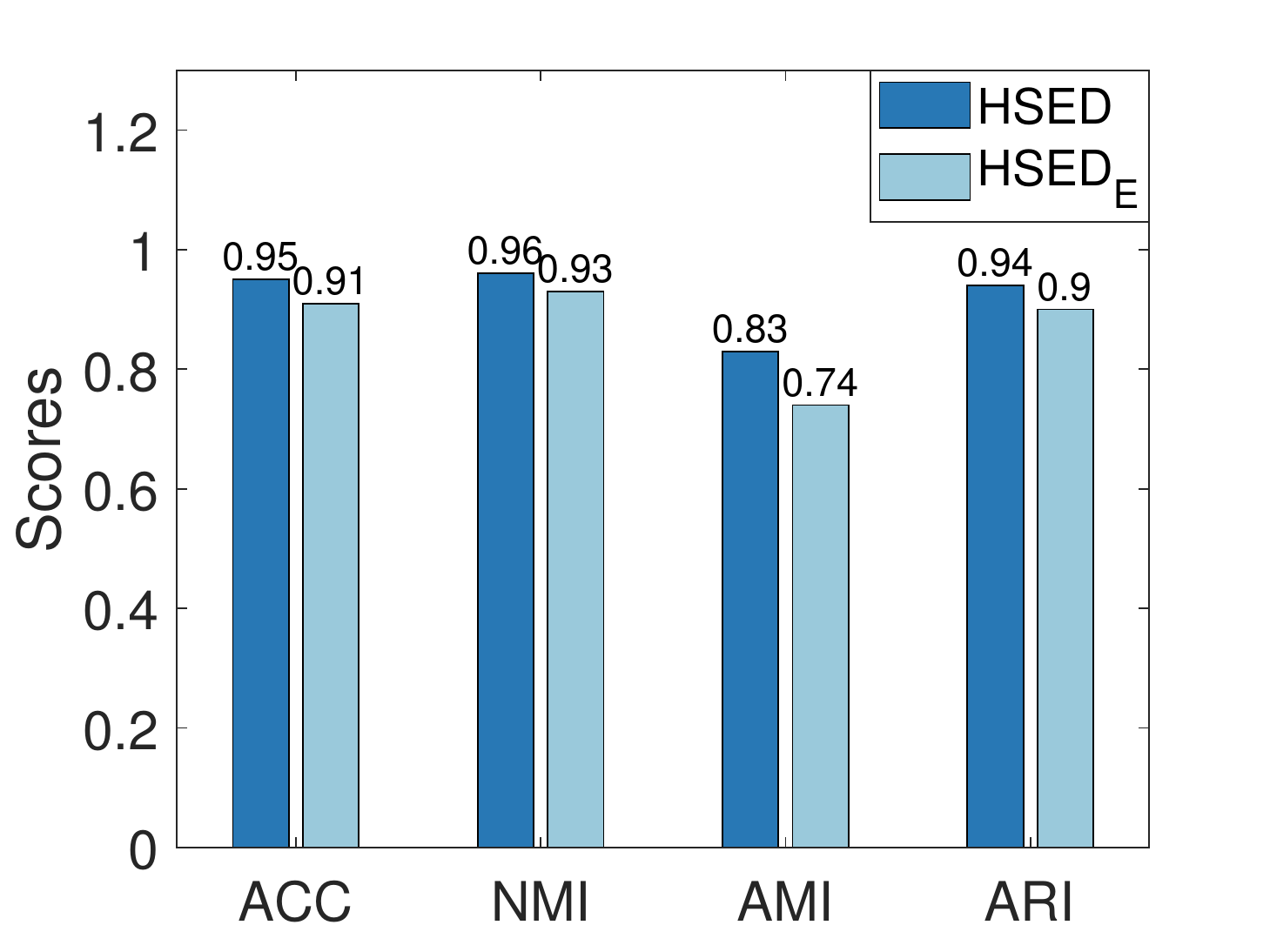}}
\subfigure[Hyperbolic space models analysis]{\includegraphics[width=0.24\textwidth]{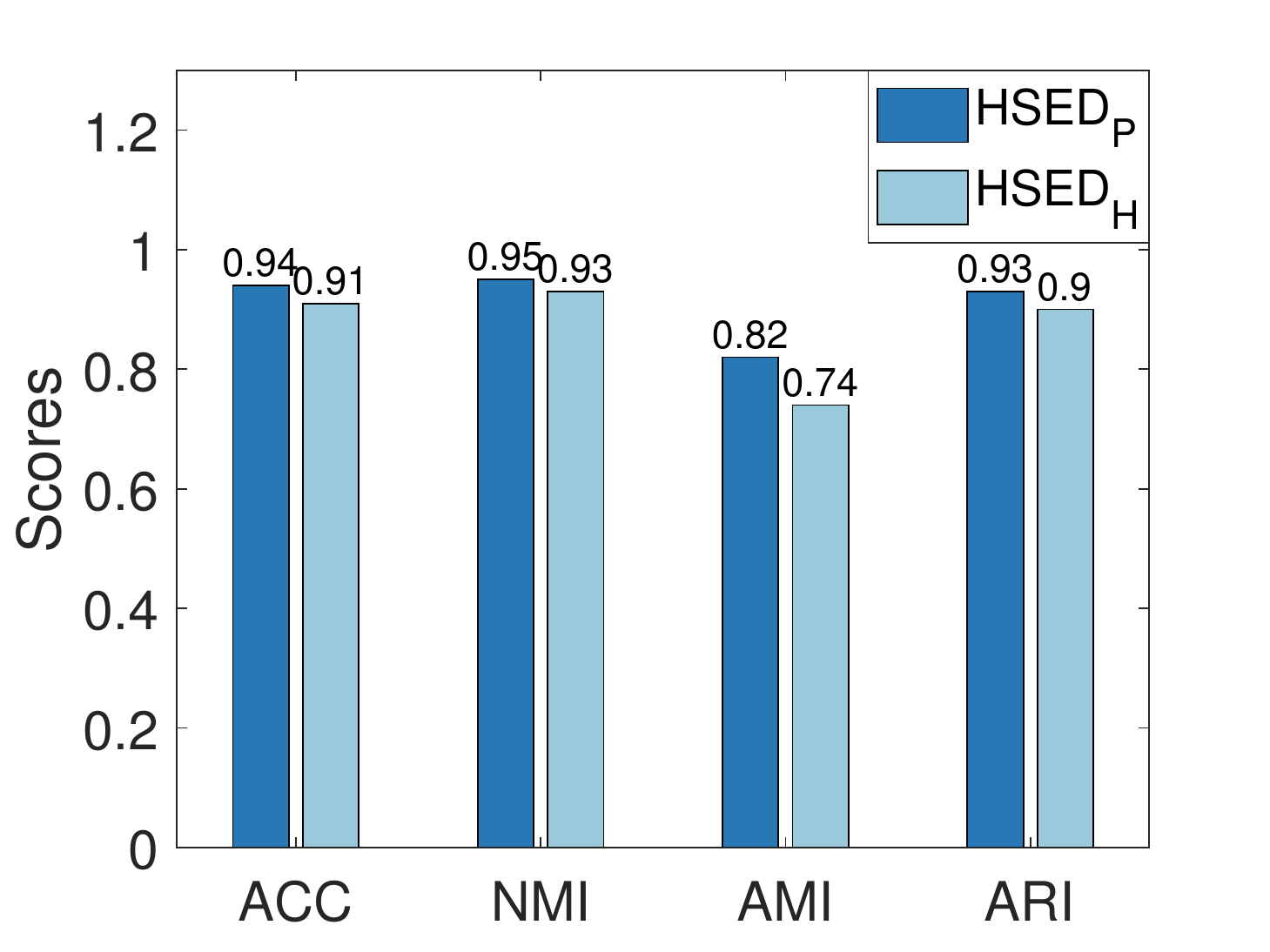}}

\caption[Hyperbolic space and their models analysis of the HSED model]{Hyperbolic space and their models analysis of the HSED model. (a) is the hyperbolic space analysis for the HSED model. “$HSED_E$” means the HSED model in Euclidean space. (b) is the hyperbolic space models' analysis for the HSED model. “$HSED_P$” means the HSED model embedding via the Poincaré ball model. “$HSED_H$” means the HSED model embedding via the Hyperboloid ball model.}
\label{fg:HSED space anaylsis}
\end{figure}

\begin{figure*}[!tb]
\centering
\subfigure[Micro-F1]{\includegraphics[width=0.24\textwidth]{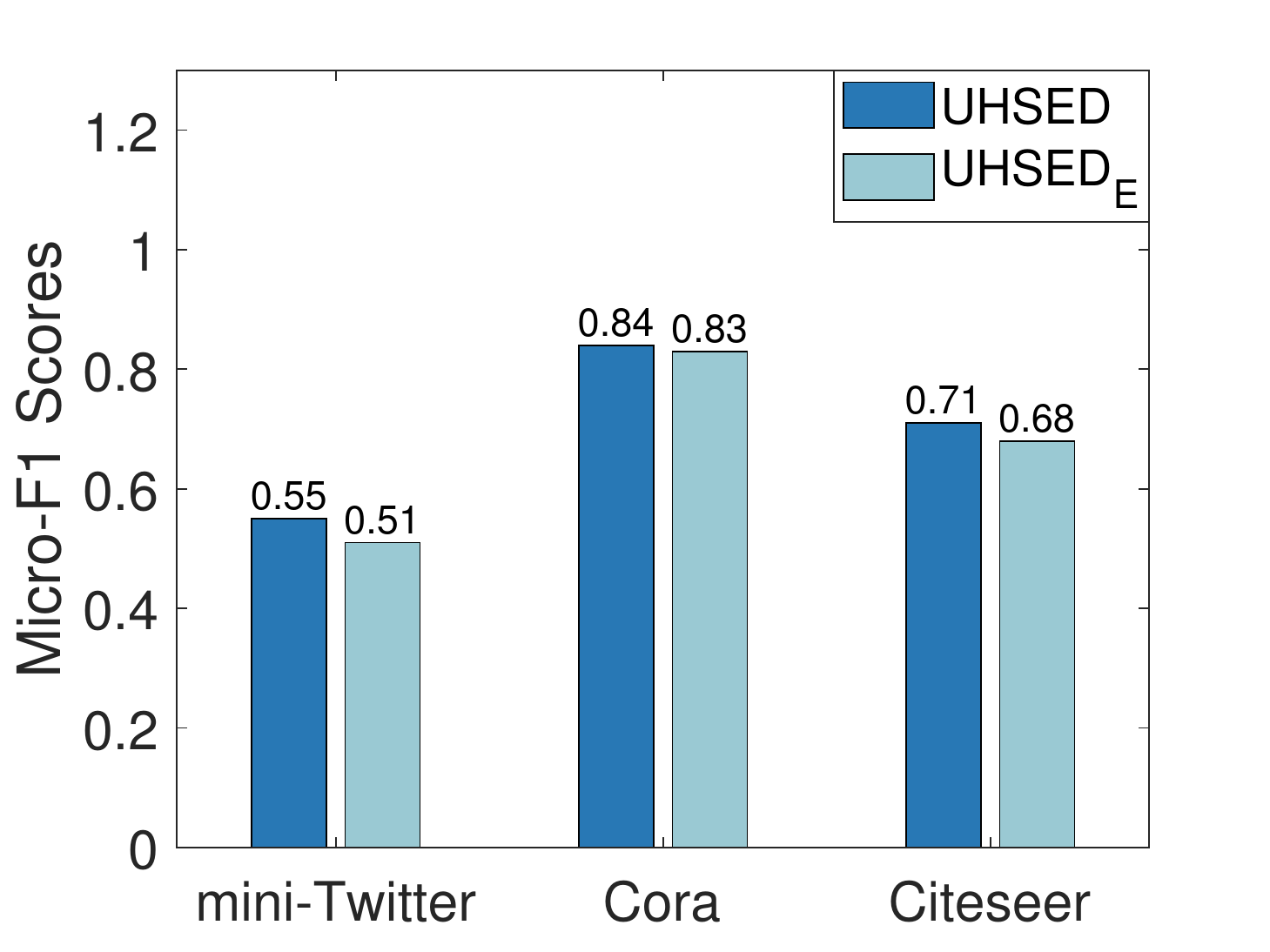}}
\subfigure[Macro-F1]{\includegraphics[width=0.24\textwidth]{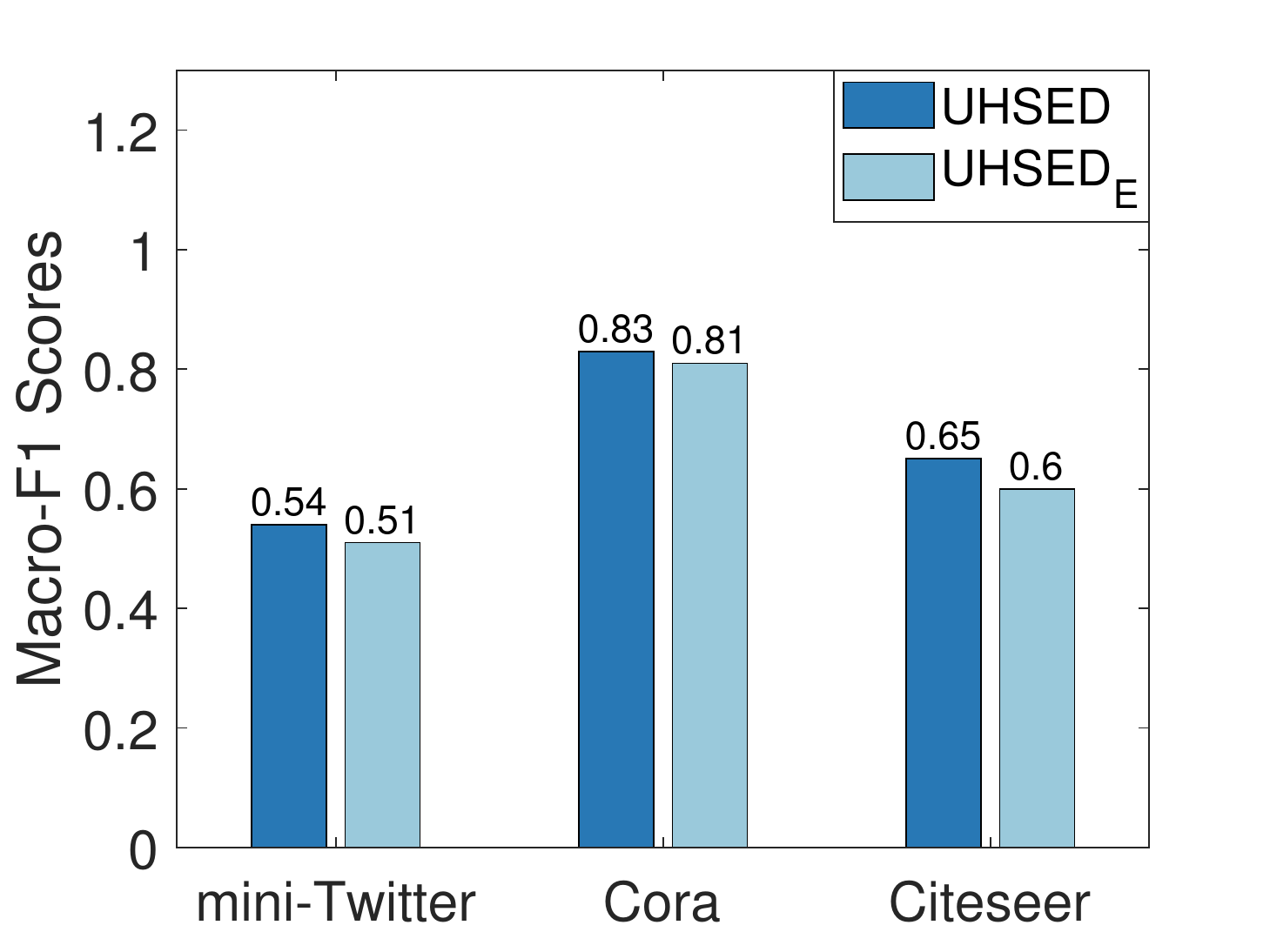}}
\subfigure[Micro-F1]{\includegraphics[width=0.24\textwidth]{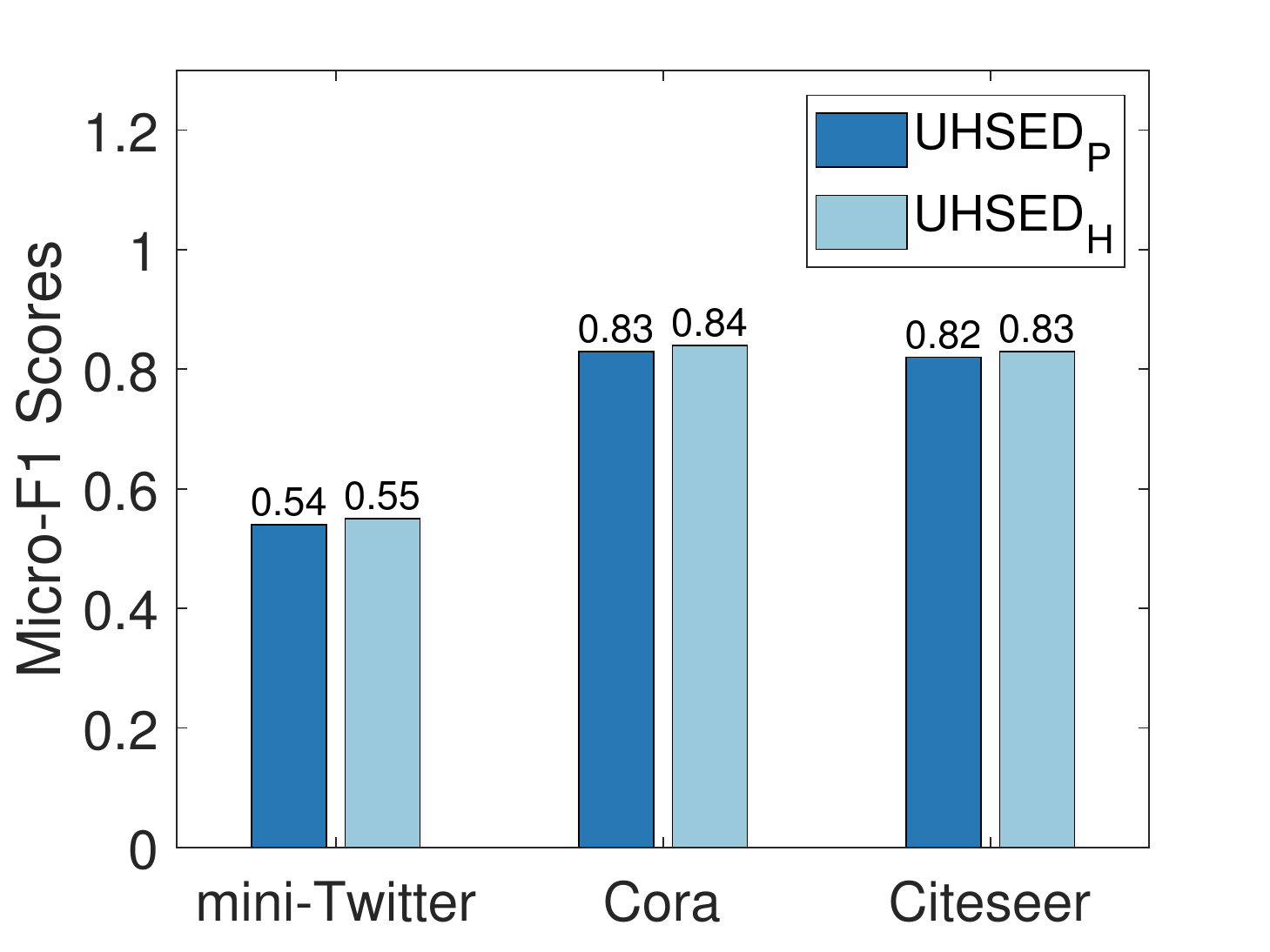}}
\subfigure[Macro-F1]{\includegraphics[width=0.24\textwidth]{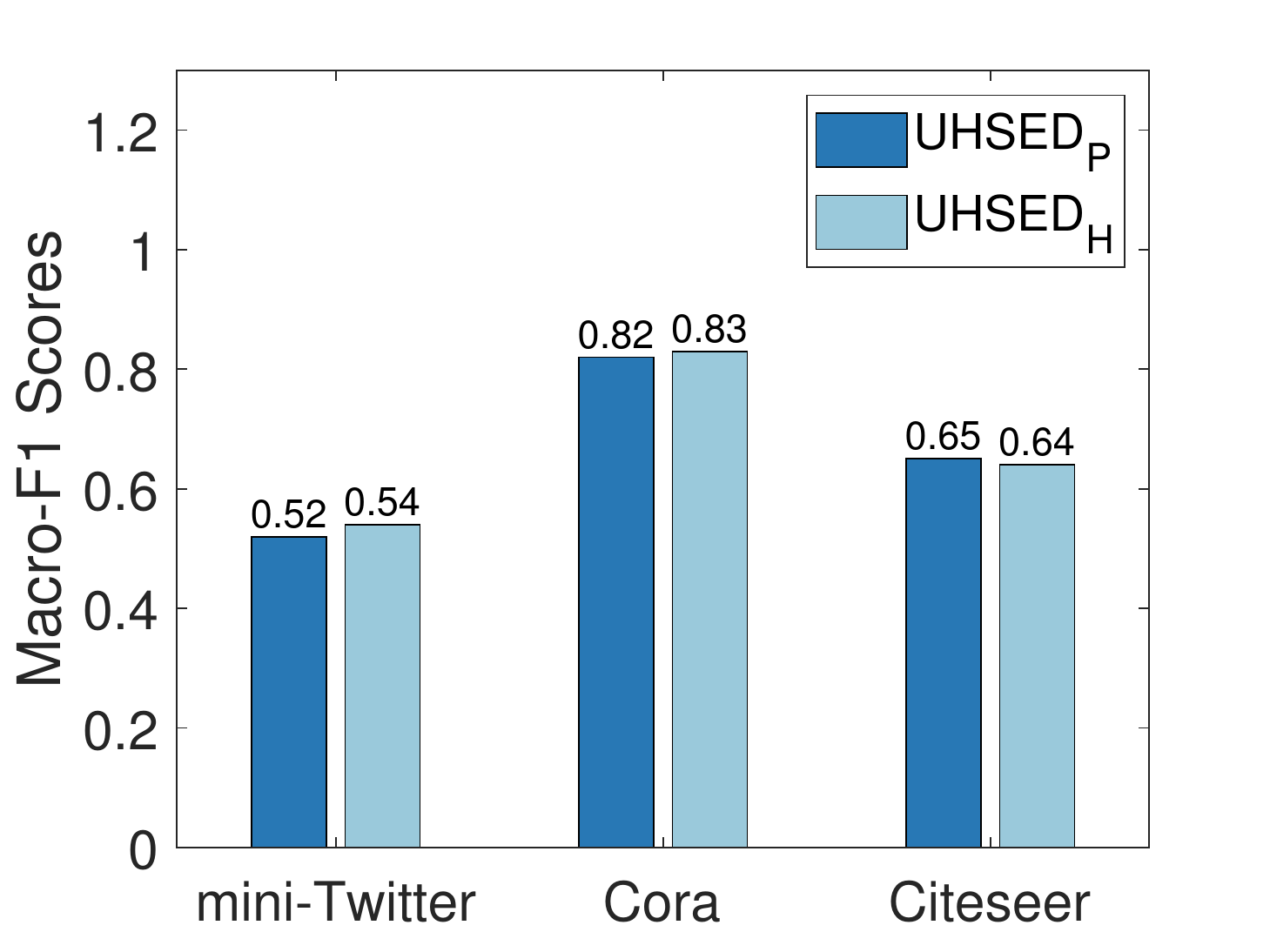}}

\caption[Hyperbolic space and their models analysis of the UHSED model]{Hyperbolic space and their models analysis of the UHSED model. (a) and (b) represent the hyperbolic space analysis for the UHSED model. “$UHSED_E$” means the UHSED model in Euclidean space. (c) and (d) represent the hyperbolic models' analysis for the UHSED model. “$UHSED_P$” means the UHSED model embedding via the Poincaré ball model. “$UHSED_H$” means the UHSED model embedding via the Hyperboloid ball model.}
\label{fg:UHSED space analysis}
\end{figure*}

To answer this question, we compared the HSED model with a variant of the HSED model that only embeds features in Euclidean space. The results are shown in Fig.  \ref{fg:HSED space anaylsis}(a). It is clear that, with the Twitter dataset, the HSED model that embeds features in hyperbolic space outperforms the Euclidean space variant of the model in all metrics, with an average improvement of $5\%$.

The results for the UHSED model are shown in Fig.  \ref{fg:UHSED space analysis}(a), (b). Here, we can see that hyperbolic space also improves the model’s performance with all datasets. Overall, the experiments clearly demonstrate the effectiveness of modelling in hyperbolic space given tree-structured data.

\subsubsection{Model Performance in Different Hyperbolic Space Models (Answer Q5)} \label{qs:hyperbolic model}

As mentioned, we applied two types of hyperbolic models – one being the Poincaré ball model, and the other being the hyperboloid model. This subsection explores the impact of these different hyperbolic models on our proposed models. The experimental results are shown in Fig. \ref{fg:HSED space anaylsis}(b) and Fig. \ref{fg:UHSED space analysis}(c), (d).

Overall, the differences between the hyperbolic models had little effect on the proposed models. However, for the HSED model, the Poincaré ball model performed better than the hyperboloid model. The reason for this is that the researcher can adjust the Poincaré ball model using gradient optimisation. Because of this, the Poincaré ball model is more suitable for representation learning.

\subsubsection{The Relationship Between Tree-like Structure and Neighbours' Aggregation (Answer Q6)} \label{qs: tree-like GCN}

Through the above experiments, we found an interesting phenomenon. The models based on GCNs do not perform well with the Twitter dataset. We speculate that this is because, as the data with a tree-like structure increases, the distance between neighbours becomes closer, which leads to a failure of the aggregation algorithm in the GCN.

\begin{table}[!htb]
    \centering
    \caption[Validating the impact of tree-structured data with GCNs]{Validating the impact of tree-structured data with GCNs (Results for the GCN-based models are marked in italic).}
    \scalebox{1.2}{
    \begin{tabular}{ccccc}
    \toprule[1 pt]
    Methods & ACC & NMI & AMI & ARI \\ \midrule
    HSED    & 0.9354 & 0.951 & 0.8189 & 0.929 \\
    MLP     & 0.8971 & 0.9207 & 0.7802 & 0.885  \\ \hline
    UHSED   & \textit{0.166} & \textit{0.3062} & \textit{0.0182} & \textit{0.0922}  \\
    GCN     & \textit{0.2552} & \textit{0.3776 }& \textit{0.0342} & \textit{0.1576 } \\
  
    \bottomrule[1pt]
    \end{tabular}
    }
    \label{tab:tree-like structure}
\end{table}

To verify our conjecture, we compared the classical MLP and GCN models with our proposed model and conducted experiments on the Twitter dataset. The results are shown in Table \ref{tab:tree-like structure}. Through these experiments, we confirmed that GCN-based models do not perform well on the Twitter dataset. Although hyperbolic space improves the models’ performance, it does not affect the model as much as the model’s structure. Therefore, we believe that tree-structured data does hinder the aggregation algorithm within GCN-based models.

\subsubsection{Time Efficiency}
\begin{figure}[!tb]
\centering
\subfigure[HSED model]{\includegraphics[width=0.24\textwidth]{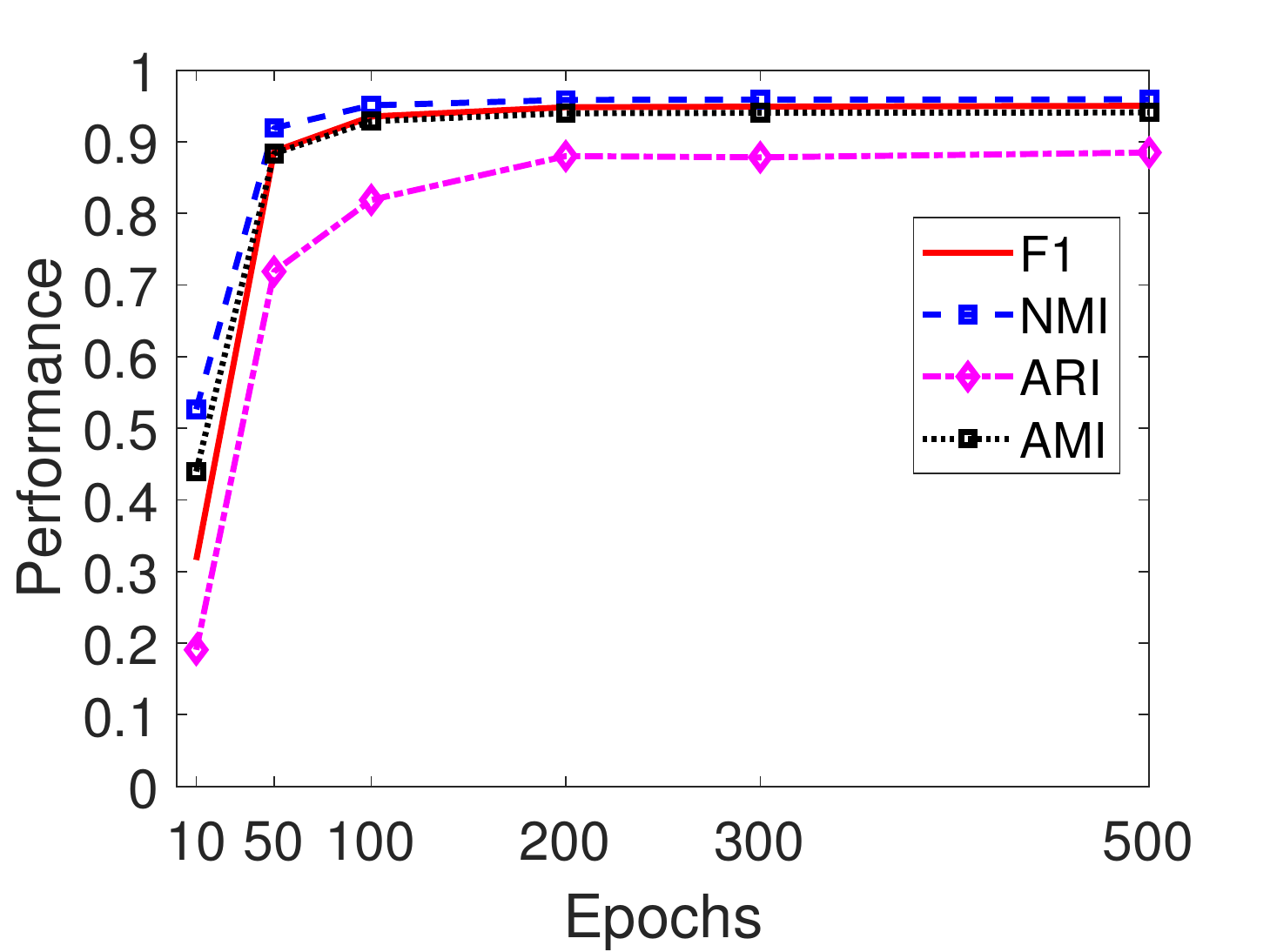}}
\subfigure[UHSED model]{\includegraphics[width=0.24\textwidth]{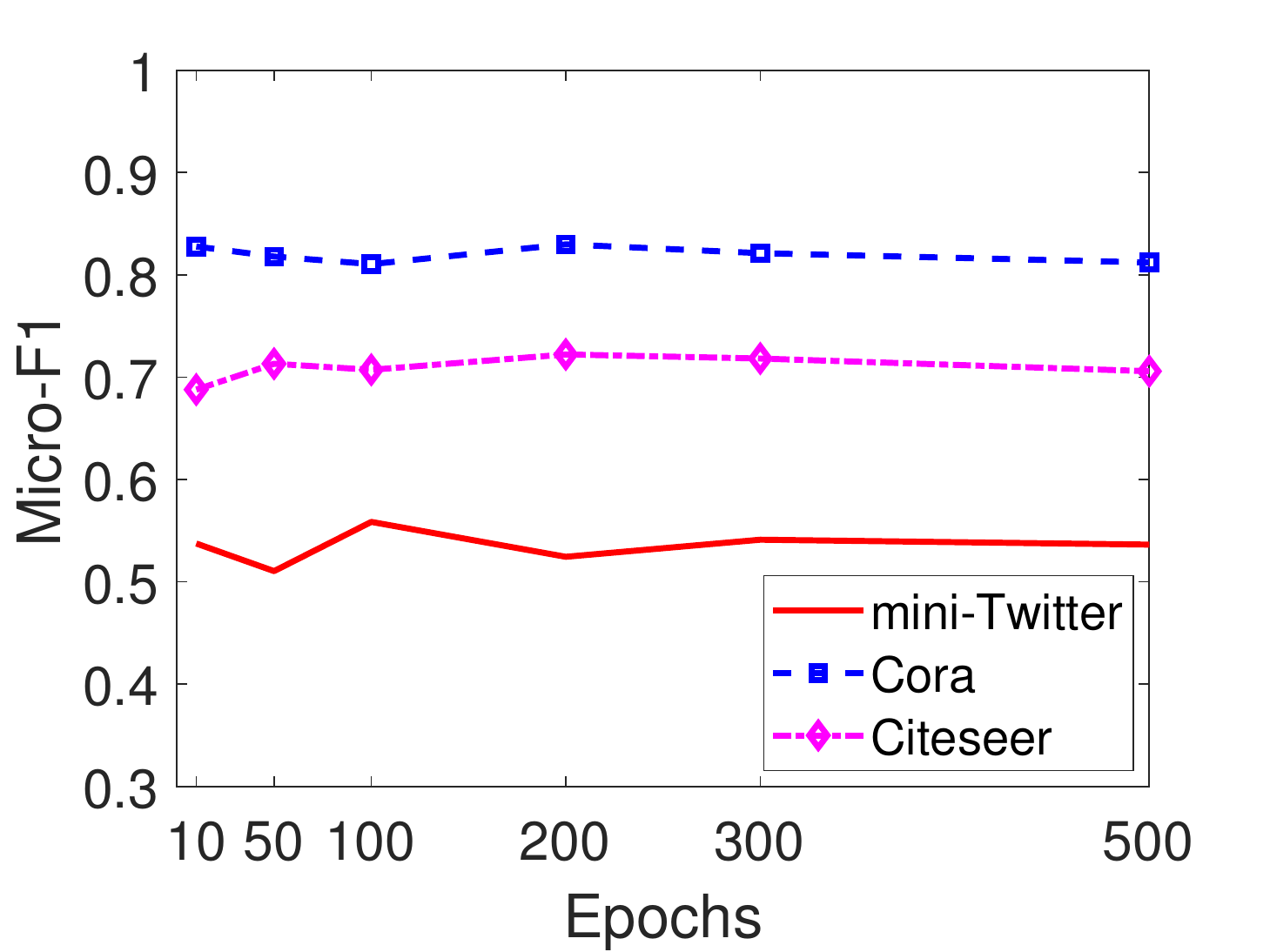}}

\caption[Effect of epoch number on model performance]{Effect of epoch number on model performance.}
\label{fg:epoch analysis}
\end{figure}

In this section, we explore the time efficiency of the baseline models and proposed models. We analyse the relationship between the number of epochs and model performance and compare the temporal performance of the baseline models and the proposed models on different datasets. The experimental results of the supervised model (HSED) and the unsupervised model (UHSED) are shown in Fig. \ref{fg:epoch analysis}, Table \ref{tab:HSED time} and Table \ref{tab:UHSED time}, respectively.

Fig. \ref{fg:epoch analysis} shows the effect of model training time on model performance. It can be seen that the performance of the model becomes better as the epoch increases. However, it is not that the larger the number of epochs, the better. We can see that for the supervised model (HSED) when the number of epochs exceeds 100, the performance of the model increases slowly. For the unsupervised model (UHSED), the number of optimal epochs varies with the dataset. Therefore, we can clearly see that increasing the training time of the model can improve its performance, but it does not mean that the longer the training time, the better the model's performance. It has a peak, and the peak varies with the dataset.

\begin{table}[!htb]
    \centering
    \caption[HSED time efficiency analysis]{HSED model time efficiency analysis. }
    \scalebox{0.79}{
    \begin{tabular}{c|cccccc|c}
    \toprule[1 pt]
    Metrics & Word2Vec & LDA & WMD & BERT  & KPGNN & FinEvent & HSED (ours) \\ \midrule
    Time & 306.1s & 77.73s & >24h & 9784s & 39.91s & >24h & 246.28s\\

    \bottomrule[1pt]
    \end{tabular}
    }
    \label{tab:HSED time}
\end{table}
Table \ref{tab:HSED time} and Table \ref{tab:UHSED time} show the time efficiency experiments of different models. For the supervised model (HSED), we can see that its time efficiency is better than most of the baseline models. The reason may be that some baseline models are not designed to be applied to big data strategies. Like WMD model needs to compare the similarity between each message before classifying the messages. It becomes very time-consuming when encountering larger datasets. The FinEvent model, on the other hand, combines data processing and training together, which will consume a lot of time in data processing when encountering large data sets, resulting in huge overall time consumption.

\begin{table}[!htb]
    \centering
    \caption[UHSED time efficiency analysis]{UHSED model time efficiency analysis. }
    \begin{tabular}{c|cc|c}
    \toprule[1 pt]
    Datasets     & DGI     & GraphGL & UHSED (ours) \\ \midrule
    mini-Twitter & 58.94s  & 112.01s & 225.6s  \\
    Cora         & 205.25s & 86.43s  & 96.2s  \\
    Citeseer     & 680.29s & 110.19s & 370.4s  \\

    \bottomrule[1pt]
    \end{tabular}
    \label{tab:UHSED time}
\end{table}

For the unsupervised model (UHSED), its time efficiency is not much different from other baseline models for different datasets. It may be that they both adopt the same graph contrastive learning framework. From this point of view, the application of hyperbolic space has little effect on time efficiency, and the change in time is more likely due to the size of the data set and the distribution mechanism of the data. Overall, our proposed hyperbolic space-based models improve the performance of social event detection without sacrificing time efficiency.

\section{Conclusion}
We proposed a Hyperbolic Social Event Detection model HSED for detecting social events in heterogenous social networks where the data is labelled, along with an unsupervised variant of the same model, UHSED, for cases where the social media data is unlabelled. This unsupervised model works around the high cost of labelling social media data by using graph contrastive learning to free the researcher from a dependency on labels. Both models preserve the rich semantic and structural information associated with heterogeneous social networks by first processing the data with Word2Vec, transforming the heterogeneous social network into a homogeneous message graph. Then, unlike other social event detection models, the models focus on the tree-like structure of social media data, for the first time employing hyperbolic space instead of Euclidean space to capture valuable semantic and structural information.  Experiments demonstrate the superiority of the proposed models for detecting social events in an offline fashion.

Notably, our experiments found that large-scale tree-structured data hinders the neighbour aggregation functions in a GCN. Moreover, the UHSED model does not perform as well as the HSED model. However, given that offline social event detection is the basis of online social event detection, we plan to shift the basis of our unsupervised model from one based on aggregation functions to one based on an MLP in future work. Furthermore, we will also consider how to apply our model to dynamic social event detection.

\ifCLASSOPTIONcompsoc
  \section*{Acknowledgments}
\else
  \section*{Acknowledgment}
\fi

This work was supported by the Australian Research Council  Projects Nos. DE200100964, LP210301259, and DP230100899.

\ifCLASSOPTIONcaptionsoff
  \newpage
\fi

\bibliographystyle{IEEEtran}
\bibliography{main}
%



%

\begin{IEEEbiography}[{\includegraphics[width=1in,height=1.25in,clip,keepaspectratio]{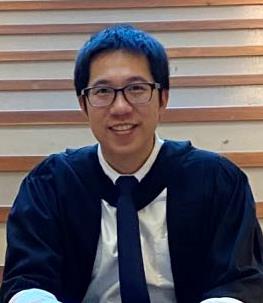}}]{Zitai Qiu}
is currently a Master of Research student at the School of Computing, Macquarie University, Sydney, Australia. He got his Master Degree from the University of Queensland, Australia. His research interests mainly include: data mining; deep learning; social event detection and machine learning.
\end{IEEEbiography}

\vspace{-1cm}
\begin{IEEEbiography}[{\includegraphics[width=1in,height=1.25in,clip,keepaspectratio]{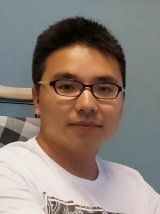}}]{Jia Wu} (M'16) is currently the Research Director for the Centre for Applied Artificial Intelligence and the Director of HDR (Higher Degree Research) in the School of Computing at Macquarie University, Sydney, Australia. 

Dr Wu received his Ph.D. degree in computer science from the University of Technology Sydney, Australia. His current research interests include data mining and machine learning. Since 2009, he has published 100+ refereed journal and conference papers, including TPAMI, TKDE, TKDD, TNNLS, TMM, KDD, ICDM, WWW, and NeurIPS.

Dr Wu has been serving as the Programme Committee Chair/Contest Chair/Publicity Chair/(Senior) Programme Committees for the prestigious data mining and artificial intelligence conferences for over 10 years, such as KDD, ICDM, WSDM, IJCAI, AAAI, WWW, NIPS, CIKM, SDM, etc. His research team was the recipient of the CIKM'22 Best Paper Runner-Up Award, ICDM'21 Best Student Paper Award, SDM'18 Best Paper Award in Data Science Track, IJCNN'17 Best Student Paper Award, and ICDM'14 Best Paper Candidate Award. Dr Wu is the Associate Editor of ACM {Transactions on Knowledge Discovery from Data} (TKDD) and Neural Networks. Dr Wu is a Senior Member of the IEEE.
\end{IEEEbiography}

\vspace{-1cm}

\begin{IEEEbiography}[{\includegraphics[width=1in,height=1.25in,clip,keepaspectratio]{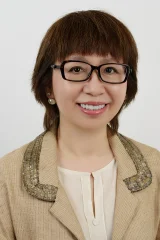}}]{Jian Yang} is a full professor at the School of Computing, Macquarie University. She received her PhD in Data Integration from the Australian National University in 1995. Her main research interests are: business process management; data science; social networks. Prof. Yang has published more than 200 journal and conference papers in international journals and conferences such as IEEE Transactions, Information Systems, Data and Knowledge Engineering, VLDB, ICDE, ICDM, CIKM, etc. She is currently serving as an Executive Committee for the Computing Research and Education Association of Australia.
\end{IEEEbiography}

\vspace{-1cm}

\begin{IEEEbiography}[{\includegraphics[width=1in,height=1.25in,clip,keepaspectratio]{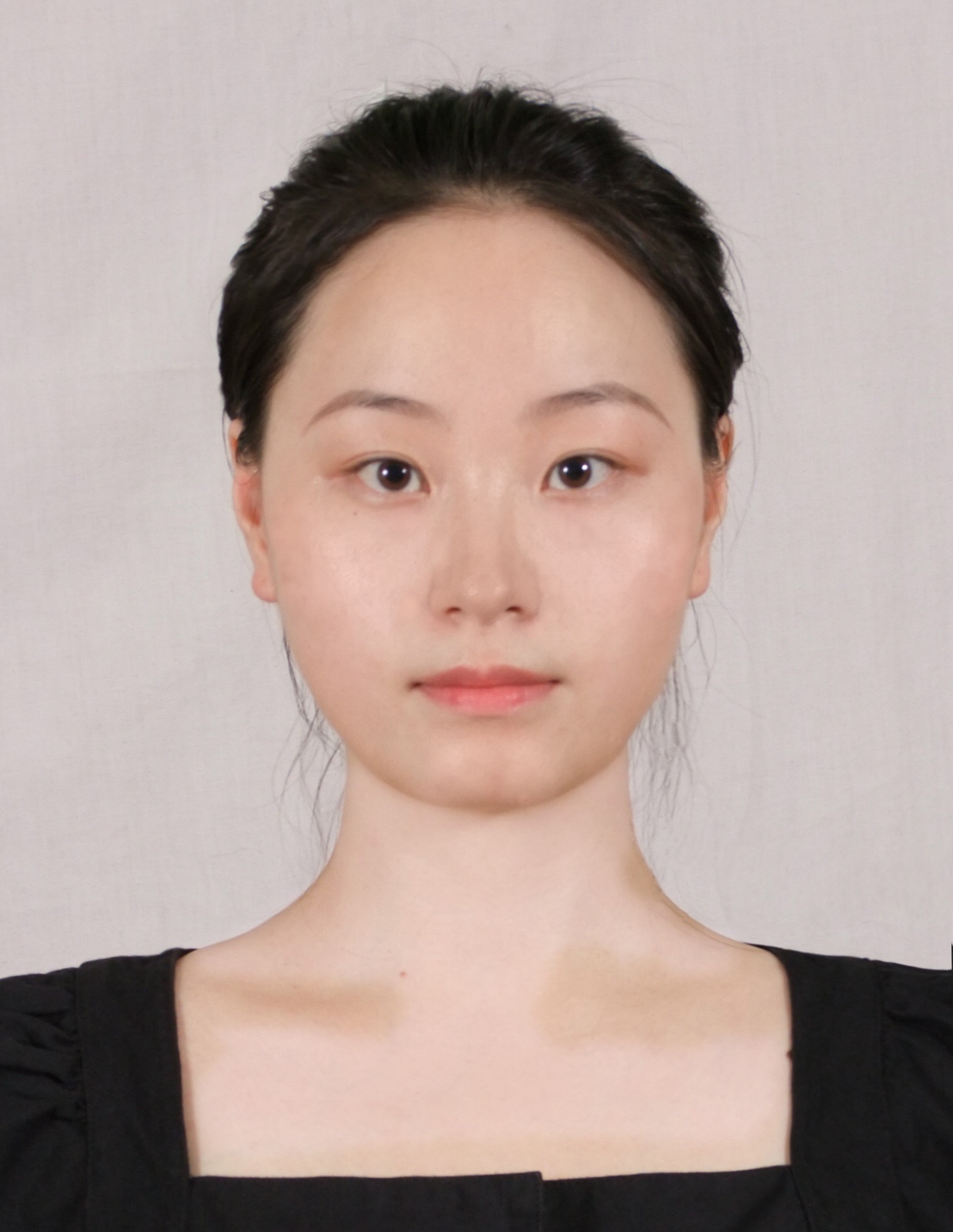}}]{Xing Su} received her M.Eng. degree in computer technology from Lanzhou University, China in 2020. She is currently a Ph.D. candidate in School of Computing at Macquarie University, Australia. Her current research interests include misinformation detection, community detection, deep learning, and social network analysis.
\end{IEEEbiography}

\vspace{-1cm}

\begin{IEEEbiography}[{\includegraphics[width=1in,height=1.25in,clip,keepaspectratio]{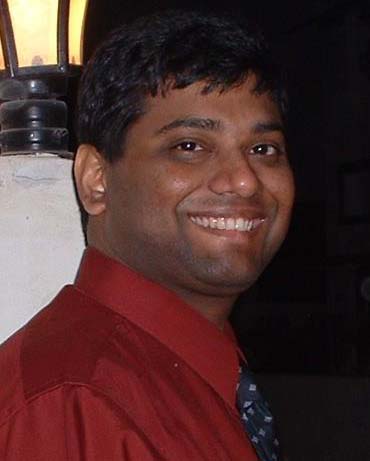}}]{Charu Aggarwal}(F'10) is a Distinguished Research Staff Member (DRSM) at the IBM T. J. Watson Research Center in Yorktown Heights, New York. He received the BS degree from IIT Kanpur, in 1993, and the PhD degree from the
Massachusetts Institute of Technology, in 1996.

He has since worked in the field of performance
analysis, databases, and data mining. He has
served on the program committees of most major
database/data mining conferences, and served as
program vice-chair of SDM 2007, ICDM 2007,
WWW 2009, and ICDM 2009. He served as an
associate editor of the IEEE Transactions on Knowledge and Data Engineering from 2004 to 2008. He is an associate editor of the ACM Transactions on Knowledge Discovery from Data, an action editor of Data Mining
and Knowledge Discovery, an associate editor of SIGKDD Explorations,
and an associate editor of KAIS. He is a fellow of the IEEE and the ACM.

\end{IEEEbiography}




\end{document}